%% file: ms.tex
\documentclass[iop, twocolappendix]{mn2e} 

\usepackage[pdftex]{graphicx}

\usepackage{amsmath, amsthm, amssymb}

\usepackage{natbib}
\usepackage{subfigure}
\usepackage{upgreek}
\usepackage{accents}
\usepackage{color}
\usepackage{hyperref}
\usepackage{dblfloatfix}

\input{macros.tex}

\title[Pre-Reionization X-ray Sources]{Decoding the X-ray Properties of Pre-Reionization Era Sources}
\author[J. Mirocha]{Jordan Mirocha\thanks{E-mail:
jordan.mirocha@colorado.edu} \\
Center for Astrophysics and Space Astronomy and Department of Astrophysical and Planetary Science \\ 
University of Colorado, Campus Box 389, Boulder, CO 80309}
\begin{document}
    
\pagerange{\pageref{firstpage}--\pageref{lastpage}} \pubyear{2014}
\maketitle

\begin{abstract}
Evolution in the X-ray luminosity -- star formation rate ($L_X$-SFR) relation
could provide the first evidence of a top-heavy stellar initial mass function
in the early universe, as the abundance of high-mass stars and binary systems
are both expected to increase with decreasing metallicity. The sky-averaged
(global) 21-cm signal has the potential to test this prediction via
constraints on the thermal history of the intergalactic medium, since X-rays
can most easily escape galaxies and heat gas on large scales. A significant
complication in the interpretation of upcoming 21-cm measurements is the
unknown spectrum of accreting black holes at high-$z$, which depends on the
mass of accreting objects and poorly constrained processes such as how
accretion disk photons are processed by the disk atmosphere and host galaxy
interstellar medium. Using a novel approach to solving the cosmological
radiative transfer equation (RTE), we show that reasonable changes in the
characteristic BH mass affects the amplitude of the 21-cm signal's minimum at
the $\sim 10-20$ mK level --- comparable to errors induced by commonly used
approximations to the RTE --- while modifications to the intrinsic disk
spectrum due to Compton scattering (bound-free absorption) can shift the
position of the minimum of the global signal by $\Delta z \approx 0.5$
($\Delta z \approx 2$), and modify its amplitude by up to $\approx 10$ mK
($\approx 50$ mK) for a given accretion history. Such deviations are larger
than the uncertainties expected of current global 21-cm signal extraction
algorithms, and could easily be confused with evolution in the $L_X$-SFR
relation.
\end{abstract}


\section{INTRODUCTION} \label{sec:Introduction}
The universe's transition from a cold and mostly neutral state after
cosmological recombination, to a hot, ionized plasma $\sim 1$ billion years
later, encodes information about the first generations of stars, galaxies, and
compact objects \citep{Barkana2001, Bromm2009}. However, two major
astrophysical milestones are likely to occur well before this Epoch of
Reionization (EoR) began in earnest, which are both valuable probes of the
high redshift universe: (1) decoupling of the excitation temperature of
ambient intergalactic hydrogen gas from the cosmic microwave background (CMB)
temperature by a soft ultraviolet background
\citep{Madau1997,Barkana2005,Wouthuysen1952,Field1958}, and (2) X-ray heating
of the intergalactic medium (IGM), eventually to temperatures above the CMB
temperature \citep{Venkatesan2001,Ricotti2004, Madau2004, Chen2004,
Ciardi2010,McQuinn2012b}. These events are expected to be visible in
absorption against the CMB at low radio frequencies, $\nu = \nu_0 (1 + z)$,
where $\nu_0 = 1420 \ \mathrm{MHz}$ is the rest frequency of the ground state
hyperfine 21-cm transition of neutral hydrogen, and $z$ is the redshift
\citep[for a review, see][]{FurlanettoOhBriggs2006}.

Studies of the pre-reionization epoch via the redshifted 21-cm line in
absorption have the potential to provide the first contemporaneous constraints
on the properties of the first stars and black holes (BHs), whose formation
channels may be fundamentally different than those of their counterparts in
the local universe \citep[e.g.,][]{Bromm1999,Abel2002b,Begelman2006}. Their
existence could dramatically alter the conditions for subsequent star and
black hole formation in their host halos, and perhaps globally, through strong
photo-dissociating and photo-ionizing radiation \citep[e.g.,][]{Haiman2000,
Kuhlen2005, Mesinger2009a, Tanaka2012, WolcottGreen2012, Jeon2014}.

In this work, we focus on the minimum of the global 21-cm signal and how its
position could be used to probe the properties of accreting BHs in the early
universe. The 21-cm minimum is well known as an indicator of heating
\citep[e.g.,][]{Furlanetto2006, Pritchard2007, Mirabel2011}, and from its
position one can obtain model-independent limits on the instantaneous heating
rate density and cumulative heating in the IGM over time \citep{Mirocha2013}.
The 21-cm maximum is also a probe of the IGM thermal history
\citep[e.g.,][]{Ripamonti2008}, though because it likely overlaps with the
early stages of reionization, one must obtain an independent measurement on
the ionization history in order to constrain the IGM temperature and heating
rate density \citep{Mirocha2013}. In either case, extracting the properties of
the heat sources themselves from the 21-cm signal is fraught with uncertainty
since the number density of X-ray sources and their individual luminosities
cannot be constrained independently by volume-averaged measures like the global 21-cm signal.

Despite such degeneracies among model parameters, accurate enough measurements
could still rule out vast expanses of a currently wide-open parameter space.
What remains could be visualized as a two-dimensional posterior probability
distribution that characterizes the likelihood that any given pair of model
parameters is correct, having marginalized over uncertainties in all
additional parameters. Two likely axes in such analyses include (1) the
characteristic mass (or virial temperature) of star-forming haloes and (2) the
X-ray luminosity per unit star formation. However, a third, and often ignored
axis that will manifest itself in such posterior probability spaces is the
spectral energy distribution (SED) of X-ray sources. The reason for this
expectation is simple: soft X-ray sources will heat the IGM more efficiently
than hard X-ray sources (at fixed total X-ray luminosity) due to the strong
frequency dependence of the bound-free absorption cross section ($\sigma
\propto \nu^{-3}$ approximately).

High-mass X-ray binaries (HMXBs) are often assumed to be the dominant source
of X-rays in models of high-$z$ galaxies. This choice is motivated by X-ray
observations of nearby star-forming galaxies \citep[see review
by][]{Fabbiano2006}, as well as theoretical models of stellar evolution, which
predict the formation of more massive stellar remnants and more binaries in
metal-poor environments \citep[e.g.,][]{Belczynski2008,Linden2010,
Mapelli2010}. Indeed, observations of star-forming galaxies are consistent
with a boost in high-mass X-ray binary populations (per unit SFR) in galaxies
out to $z \sim 4-6$ \citep{BasuZych2013,Kaaret2014}, as is the unresolved
fraction of the cosmic X-ray background \citep{Dijkstra2012}.
Though direct constraints on the $z \gtrsim 4$ population are weak, local
analogs of high-$z$ galaxies exhibit a factor of $\sim 10$ enhancement in the
normalization of the X-ray luminosity function (XLF) in metal-poor galaxies
relative to galaxies with $\sim$ solar metallicity
\citep[e.g.][]{Kaaret2011,Prestwich2013,Brorby2014}.

Even if HMXBs are the dominant sources of X-rays in the early universe, there
are various remaining uncertainties that may affect the global 21-cm signal
and inferences drawn from the position of its minimum. Our focus is on
modifications of the 21-cm signal brought about by variation in the
characteristic mass of accreting objects and the reprocessing of their
intrinsic emission spectrum by intervening material. Theoretical
investigations of this sort can provide vital information to upcoming 21-cm
experiments that seek to detect the absorption trough, such as the Dark Ages
Radio Explorer \citep[DARE;][]{Burns2012}, the Large Aperature Experiment to
Detect the Dark Ages \citep[LEDA;][]{Greenhill2012}, and the SCI-HI experiment
\citep{Voytek2014}. For instance, how accurately must the 21-cm absorption
trough be measured in order to distinguish models for the first X-ray sources?

The challenge for such studies is solving the cosmological radiative transfer
equation (RTE) in a way that 1) accurately couples the radiation field from
sources to the thermal and ionization state of the IGM, and 2) does so quickly
enough that a large volume of parameter space may be surveyed. Recent studies
have taken the first steps toward this goal by identifying SEDs likely to be
representative of high-$z$ sources \citep[e.g.,][]{Power2013}. Some have
applied semi-numeric schemes to predict how these SEDs contribute to the
ionizing background \citep{Power2013,Fragos2013}, while others have studied
the influence of realistic X-ray SEDs on the sky-averaged 21-cm signal and the
21-cm power spectrum \citep{Ripamonti2008,Fialkov2014}. Our focus is
complementary: rather than calculating the ionizing background strength or
21-cm signal that arise using ``best guess'' inputs for the SED of X-ray
sources, we quantify how reasonable deviations from best guess SEDs can
complicate inferences drawn from the signal.

The outline of this paper is as follows. In Section 2, we introduce our framework for cosmological radiative transfer and the global 21-cm signal. In Section 3, we describe our implementation of the \citet{Haardt1996} method for discretizing the RTE and test its capabilities. In Section 4, we use this scheme to investigate the impact of SED variations on the global 21-cm signal. Discussion and conclusions are in Sections 5 and 6, respectively. We adopt WMAP7+BAO+SNIa  cosmological parameters ($\Omega_{\Lambda,0} = 0.728$, $\Omega_{b,0}=0.044$, $H_0=70.2 \ \text{km} \ \text{s}^{-1} \ \text{Mpc}^{-1}$, $\sigma_8=0.807$, $n=0.96$) throughout \citep{Komatsu2011}.

\section{THEORETICAL FRAMEWORK}
As in \citet{Furlanetto2006}, we divide the IGM into two components: 1) the ``bulk IGM,'' which is mostly neutral and thus capable of producing a 21-cm signature, and 2) HII regions, which are fully ionized and thus dark at redshifted 21-cm wavelengths. This approach is expected to break down in the late stages of reionization when the distinction between HII regions and the ``neutral'' IGM becomes less clear. However, our focus in this paper is on the pre-reionization era so we expect this formalism to be reasonably accurate. 

There are three key steps one must take in order to generate a synthetic global 21-cm signal within this framework. Starting from a model for the volume-averaged emissivity of astrophysical sources, which we denote as $\upepsilon_{\nu}(z)$ or $\enu(z)$, further subdivided into a bolometric luminosity density (as a function of redshift) and SED (could also evolve with redshift in general), one must
\begin{enumerate}
    \item Determine the mean radiation background pervading the space between galaxies (the so-called ``metagalactic'' radiation background), including the effects of geometrical dilution, redshifting, and bound-free absorption by neutral gas in the IGM. We denote this angle-averaged background radiation intensity as $J_{\nu}$ or $\Jhatnu$.
    \item Once the background intensity is in hand, compute the ionization rate density, $\Gamma_{\HI}$, and heating rate density, $\eX$, in the bulk IGM. 
    \item Given the ionization and heating rate densities, we can then solve for the rate of change in the ionized fraction, $x_e$, and temperature, $T_K$, of the bulk IGM gas. The rate of change in the volume filling fraction of HII regions, $x_i$, is related more simply to the rate of baryonic collapse in haloes above a fixed virial temperature, $\Tmin$, at the redshift of interest.
\end{enumerate}
Once the thermal and ionization state of the IGM and the background intensity at the $\Lya$ resonance are known, a 21-cm signal can be computed. In this section, we will go through each of these steps in turn.

\subsection{Astrophysical Models}
We assume throughout that the volume-averaged emissivity is proportional to
the rate of collapse, $\enu(z) \propto d\fcoll/dt$, where \begin{equation}
\fcoll = \rho_m^{-1}(z) \int_{m_{\min}}^{\infty} m n(m) dm \end{equation} is
the fraction of gas in collapsed haloes more massive than $m_{\min}$. Here,
$\rho_m(z)$ is the mean co-moving mass density of the universe and $n(m) dm$
is the co-moving number density of haloes with masses in the range $(m,
m+dm)$. We compute $n(m)$ using the \texttt{hmf-calc} code \citep{Murray2013},
which depends on the \textit{Code for Anisotropies in the Microwave
Background} \citep[CAMB;][]{Lewis2000}. We choose a fixed minimum virial
temperature $\Tmin \geq 10^4 \ \text{K}$ corresponding to the atomic cooling
threshold \citep[Eq. 26;][]{Barkana2001}, which imposes redshift evolution in
$m_{\min}$.

Our model for the emissivity is then
\begin{equation}
    \upepsilon_{\nu}(z) = \rhobbar c_i f_i \frac{d\fcoll}{dt} I_{\nu}, \label{eq:GeneralizedEmissivity}
\end{equation}
where $\rhobbar$ is the mean baryon density today, $c_i$ is a physically (or observationally) motivated normalization factor that converts baryonic collapse into energy output in some emission band $i$ (e.g., $\Lya$, soft UV, X-ray), while $f_i$ is a free parameter introduced to signify uncertainty in how $c_i$ evolves with redshift. The parameter $I_{\nu}$ represents the SED of astrophysical sources, and is normalized such that $\int I_{\nu} d\nu = 1$. We postpone a more detailed discussion of our choices for $c_i$, $I_{\nu}$, and what we mean by ``astrophysical sources'' to Section \ref{sec:Results}.

\subsection{Cosmological Radiative Transfer}
Given the volume-averaged emissivity, $\upepsilon_{\nu}$, the next step in computing the global 21-cm signal is to obtain the angle-averaged background intensity, $J_{\nu}$. To do so, one must solve the cosmological RTE,
\begin{multline}
    \left(\frac{\partial}{\partial t} - \nu \Hofz \frac{\partial}{\partial \nu} \right) J_{\nu}(z) + 3 \Hofz J_{\nu}(z) = -c \alpha_{\nu} J_{\nu}(z) \\ 
    + \frac{c}{4\pi} \upepsilon_{\nu}(z) (1 + z)^3 \label{eq:RTE_DE}
\end{multline}
where $H$ is the Hubble parameter, which we take to be $H(z) \approx H_0 \Omega_{m,0} (1+z)^{3/2}$ as is appropriate in the high-$z$ matter-dominated universe, and $c$ is the speed of light. This equation treats the IGM as an isotropic source and sink of radiation, parameterized by the co-moving volume emissivity, $\upepsilon_{\nu}$ (here in units of $\coemissivityunitsenergy$, where ``cMpc'' is short for ``co-moving Mpc''), and the absorption coefficient, $\alpha_{\nu}$, which is related to the optical depth via $d\tau_{\nu} = \alpha_{\nu} ds$, where $ds$ is a path length.  The solution is cleanly expressed if we write the flux and emissivity in units of photon number (which we denote with ``hats,'' i.e., $[\Jhatnu]=\intensityunitsnumber$ and $[\enu]=\coemissivityunitsnumber$), 
\begin{equation}
    \Jhatnu(z) = \frac{c}{4\pi} (1 + z)^2 \int_{z}^{z_f} \frac{\enuprime(z^{\prime})}{H(z^{\prime})} e^{-\overline{\tau}_{\nu}} dz^{\prime} . \label{eq:AngleAveragedFlux}
\end{equation}    
The ``first light redshift'' when astrophysical sources first turn on is denoted by $z_f$, while the emission frequency, $\nu^{\prime}$, of a photon emitted at redshift $z^{\prime}$ and observed at frequency $\nu$ and redshift $z$ is
\begin{equation}
    \nu^{\prime} = \nu \left(\frac{1 + z^{\prime}}{1 + z}\right) . \label{eq:EmissionFrequency}
\end{equation}
The optical depth is a sum over absorbing species,
\begin{equation}
    \overline{\tau}_{\nu}(z, z^{\prime}) = \sum_j \int_{z}^{z^{\prime}} n_j(z^{\dprime}) \sigma_{j, \nu^{\dprime}} \frac{dl}{dz^{\dprime}}dz^{\dprime} \label{eq:OpticalDepth}
\end{equation}
where $dl/dz = c / H(z) / (1 + z)$ is the proper cosmological line element,
and $\sigma_{j, \nu}$ is the bound-free absorption cross section of species
$j=\HI,\HeI,\HeII$ with number density $n_j$. We use the fits of
\citet{Verner1996} to compute $\sigma_{j,\nu}$ unless stated otherwise, assume
the ionized fraction of hydrogen and singly ionized helium are equal (i.e.,
$\xHII = \xHeII$), and neglect $\HeII$ entirely (i.e., $\xHeIII = 0$). We will
revisit this helium approximation in Section \ref{sec:Discussion}.

The $\Lya$ background intensity, which determines the strength of Wouthuysen-Field coupling \citep{Wouthuysen1952, Field1958}, is computed analogously via
\begin{equation}
    \widehat{J}_{\alpha}(z) = \frac{c}{4\pi} (1 + z)^2 \sum_{n = 2}^{\nmax} \frecn \int_z^{z_{\max}^{(n)}} \frac{\enuprime(z^{\prime})}{H(z^{\prime})} dz^{\prime} \label{eq:LymanAlphaFlux}
\end{equation}
where $\frecn$ is the ``recycling fraction,'' that is, the fraction of photons that redshift into a $\Lyn$ resonance that ultimately cascade through the $\Lya$ resonance \citep{Pritchard2006}. We truncate the sum over $\Lyn$ levels at $n_{\max}=23$ as in \citet{Barkana2005}, and neglect absorption by intergalactic $H_2$. The upper bound of the definite integral,
\begin{equation}
    1 + z_{\max}^{(n)} = (1 + z) \frac{\left[1 - (n + 1)^{-2}\right]}{1 - n^{-2}} ,
\end{equation}
is set by the horizon of $\Lyn$ photons -- a photon redshifting through the  $\Lyn$ resonance at $z$ could only have been emitted at $z^{\prime} < z_{\max}^{(n)}$, since emission at slightly higher redshift would mean the photon redshifted through the $\text{Ly}(n+1)$ resonance.

Our code can be used to calculate the full ``sawtooth'' modulation of the soft
UV background \citep{Haiman1997} though we ignore such effects in this work
given that our focus is on X-ray heating. Preservation of the background
spectrum in the Lyman-Werner band and at even lower photon energies is crucial
for studies of feedback, but because we have made no attempt to model $H_2$
photo-dissociation or $H^{-}$ photo-detachment, we neglect a detailed
treatment of radiative transfer at energies below $h\nu = 13.6$ eV and instead
assume a flat UV spectrum between $\Lya$ and the Lyman-limit and
``instantaneous'' emission only, such that the $\Lya$ background at any
redshift is proportional to the $\Lya$ emissivity,
$\hat{\upepsilon}_{\alpha}$, at that redshift. Similarly, the growth of HII
regions is governed by the instantaneous ionizing photon luminosity, though
more general solutions would self-consistently include a soft UV background
that arises during the EoR due to rest-frame X-ray emission from much higher
redshifts.

\subsection{Ionization \& Heating Rates}
With the background radiation intensity, $J_{\nu}$, in hand, one can compute the ionization and heating this background causes in the bulk IGM. To calculate the ionization rate density, we integrate the background intensity over frequency,
\begin{equation}
    \Gamma_{\HI}(z) = 4 \pi \nH(z) \int_{\nu_{\min}}^{\nu_{\max}} \Jhatnu \sigma_{\nu,\HI} d\nu ,
\end{equation}
where $\nH = \nHbar (1 + z)^3$ and $\nHbar$ is the number density of hydrogen atoms today. The ionization rate in the bulk IGM due to fast secondary electrons \citep[e.g.,][]{Shull1985,Furlanetto2010} is computed similarly,
\begin{equation}
    \gamma_{\HI}(z) = 4 \pi \sum_j n_j \int_{\nu_{\min}}^{\nu_{\max}} \fion \Jhatnu \sigma_{\nu,j} (h\nu - h\nu_j) \frac{d\nu}{h\nu} \label{eq:HeatingRateDensity} ,
\end{equation}
and analogously, the heating rate density,
\begin{equation}
    \eX(z) = 4 \pi \sum_j n_j \int_{\nu_{\min}}^{\nu_{\max}} \fheat \Jhatnu \sigma_{\nu,j} (h\nu - h\nu_j) d\nu \label{eq:HeatingRateDensity} ,
\end{equation}
where $h\nu_j$ is the ionization threshold energy for species $j$, with number
density $n_j$, and $\nu_{\min}$ and $\nu_{\max}$ are the minimum and maximum
frequency at which sources emit, respectively. $\fion$ and $\fheat$ are the
fractions of photo-electron energy deposited as further hydrogen ionization
and heat, respectively, which we compute using the tables of
\citet{Furlanetto2010} unless otherwise stated.

\subsection{Global 21-cm Signal} \label{sec:Global21cmSignal}
Finally, given the ionization and heating rates, $\Gamma_{\HI}$, $\gamma_{\HI}$, and $\eX$, we evolve the ionized fraction in the bulk IGM via
\begin{equation}
    \frac{d x_e}{dt} = (\Gamma_{\HI} + \gamma_{\HI}) (1 - x_e) - \alpha_{\text{B}} \nel x_e \label{eq:BulkIGM}    
\end{equation}
and the volume filling factor of HII regions, $x_i$, via
\begin{equation}
    \frac{dx_i}{dt} = \fstar \fesc \Nion \nbbar \frac{d\fcoll}{dt} (1 - x_e) - \alpha_{\text{A}} C(z) n_e x_i
\end{equation}
where $\nbbar$ is the baryon number density today, $\alpha_{\text{A}}$ and
$\alpha_{\text{B}}$ are the case-A and case-B recombination coefficients,
respectively, $\nel = \nHII + \nHeII$ is the proper number density of
electrons, $\fstar$ is the star-formation efficiency, $\fesc$ the fraction of
ionizing photons that escape their host galaxies, $\Nion$ the number of
ionizing photons emitted per baryon in star formation, and $C(z)$ is the
clumping factor. We average the ionization state of the bulk IGM and the
volume filling factor of HII regions to determine the mean ionized fraction,
i.e., $\xibar = x_i + (1 - x_i) x_e$, which dictates the IGM optical depth
(Eq. \ref{eq:OpticalDepth}). We take $C(z) = \mathrm{constant} = 1$ for
simplicity, as our focus is on the IGM thermal history, though our results are
relatively insensitive to this choice as we terminate our calculations once
the 21-cm signal reaches its emission peak, at which time the IGM is typically
only $\sim 10-20$\% ionized.

The kinetic temperature of the bulk IGM is evolved via
\begin{align}
    \frac{3}{2}\frac{d}{dt}\left(\frac{\kB T_k \ntot}{\mu}\right) & = \eX + \ecomp - \mathcal{C} \label{eq:TemperatureEvolution} 
\end{align}
where $\ecomp$ is Compton heating rate density and $\mathcal{C}$ represents all cooling processes, which we take to include Hubble cooling, collisional ionization cooling, recombination cooling, and collisional excitation cooling using the formulae provided by \citet{Fukugita1994}. Equations \ref{eq:BulkIGM}-\ref{eq:TemperatureEvolution} are solved using the radiative transfer code\footnote{\href{https://bitbucket.org/mirochaj/rt1d}{https://bitbucket.org/mirochaj/rt1d}} described in \citet{Mirocha2012}.

Given $T_K$, $x_i$, $x_e$, and $\widehat{J}_{\alpha}$, we can compute the sky-averaged 21-cm signal via \citep[e.g.,][]{Furlanetto2006}
\begin{equation}
    \dTb \simeq 27 (1 - \xibar) \left(\frac{\Obnow h^2}{0.023} \right) \left(\frac{0.15}{\Omnow h^2} \frac{1 + z}{10} \right)^{1/2} \left(1 - \frac{\Tcmb}{T_{\mathrm{S}}} \right) , \label{eq:dTb}
\end{equation}
where
\begin{equation}
    T_S^{-1} \approx \frac{T_{\gamma}^{-1} + x_c T_K^{-1} + x_{\alpha} T_{\alpha}^{-1}}{1 + x_c + x_{\alpha}}
\end{equation}
is the excitation or ``spin'' temperature of neutral hydrogen, which
characterizes the number of hydrogen atoms in the hyperfine triplet state
relative to the singlet state, and $T_{\alpha} \simeq T_K$. We compute the
collisional coupling coefficient using the tabulated values in
\citet{Zygelman2005}, and take $x_{\alpha} = 1.81 \times 10^{11}
\widehat{J}_{\alpha} / (1. + z)$, i.e., we ignore detailed line profile
effects \citep{Chen2004,FurlanettoPritchard2006, Chuzhoy2006,Hirata2006}.

\section{The Code} \label{sec:Methods}
The first step in our procedure for computing the global 21-cm signal --
determining the background radiation intensity -- is the most difficult. This
step is often treated approximately, by truncating the integration limits in
Equations \ref{eq:AngleAveragedFlux} (for $J_{\nu}$) and
\ref{eq:HeatingRateDensity} (for $\eX$) \citep[e.g.,][]{Mesinger2011}, or
neglected entirely \citep[e.g.,][]{Furlanetto2006} in the interest of speed.
In what follows, we will show that doing so can lead to large errors in the
global 21-cm signal, but more importantly, such approaches preclude detailed
studies of SED effects.

Other recent works guide the reader through Equations
\ref{eq:AngleAveragedFlux} and \ref{eq:HeatingRateDensity}, but give few
details about how the equations are solved numerically
\citep[e.g.,][]{Pritchard2007,Santos2010,Tanaka2012}. Brute-force solutions to
Equation \ref{eq:HeatingRateDensity} are accurate but extremely expensive,
while seemingly innocuous discretization schemes introduced for speed can
induce errors in the global 21-cm comparable in magnitude to several physical
effects we consider in Section \ref{sec:Results}. The goal of this Section is
to forestall confusion about our methods, and to examine the computational
expense of solving Equation \ref{eq:HeatingRateDensity} accurately.

\subsection{Discretizing the Radiative Transfer Equation} \label{sec:Discretization}
Obtaining precise solutions to Equation (\ref{eq:AngleAveragedFlux}) is difficult because the integrand is expensive to calculate, mostly due to the optical depth term, which is itself an integral function (Equation \ref{eq:OpticalDepth}). One approach that limits the number of times the integrand in Equation (\ref{eq:AngleAveragedFlux}) must be evaluated is to discretize in redshift and frequency, and tabulate the optical depth \textit{a-priori}. Care must be taken, however, as under-sampling the optical depth can lead to large errors in the background radiation intensity. This technique also requires one to assume an ionization history \textit{a-priori}, $\xibar(z)$, which we take to be $\xibar(z) = \text{constant} = 0$ over the redshift interval $10 \leq z \leq 40$. We defer a detailed discussion of this assumption to Section \ref{sec:Discussion}. 

\begin{figure}
\includegraphics[width=0.49\textwidth]{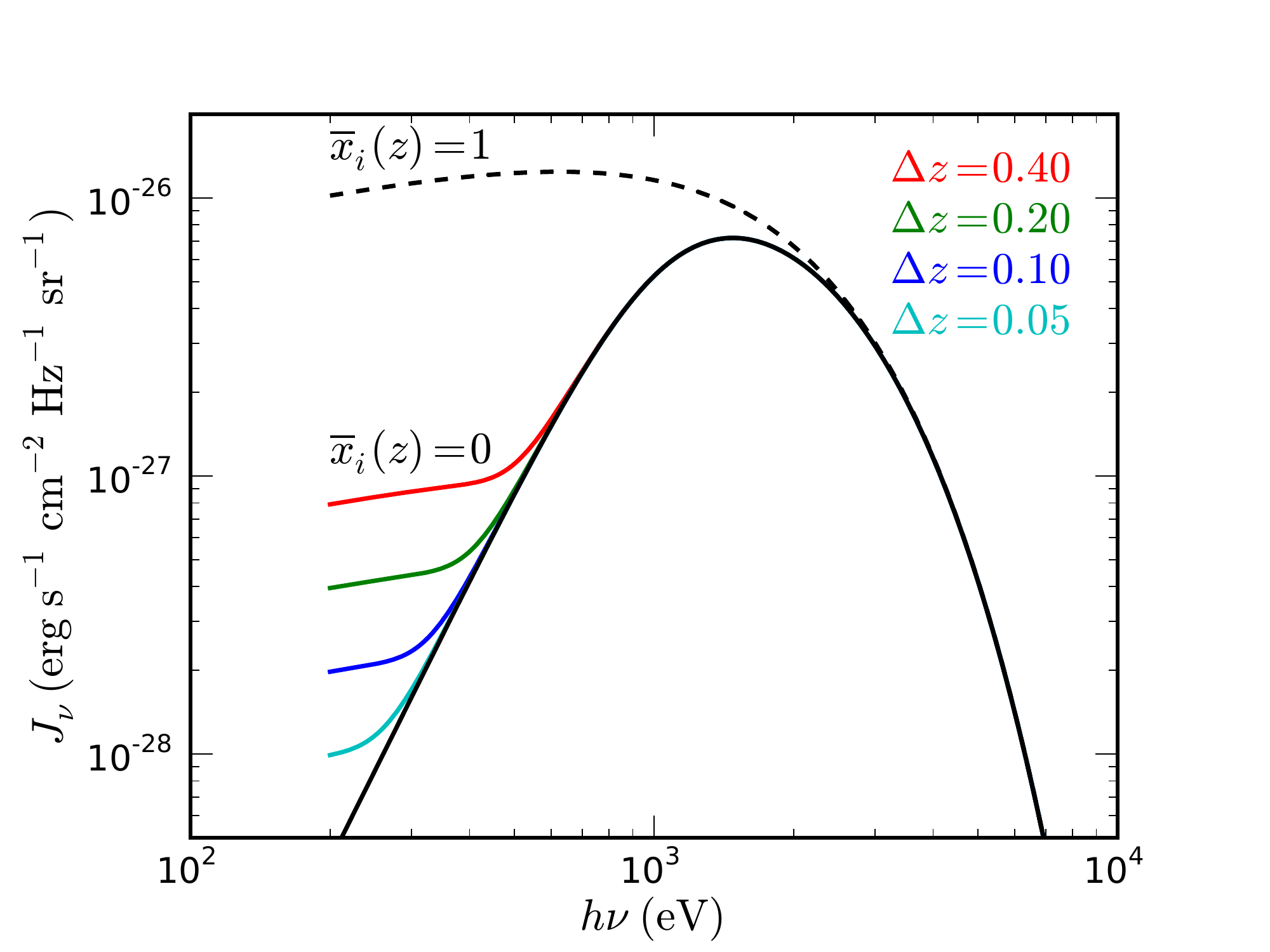}
\caption{X-ray background intensity, $J_{\nu}$, at $z=20$ assuming a population of $10 \ \Msun$ BHs. The IGM optical depth, $\taubar$, is sampled at  128 logarithmically spaced frequencies between $0.2$ and $30$ keV, and linearly in redshift by $\Delta z = 0.4$ (red), $0.2$ (green), $0.1$ (blue), and $0.05$ (cyan). Poor redshift resolution always leads to overestimates of the background intensity at soft X-ray energies ($h\nu \lesssim 0.5$ keV) since the integrand is a rapidly evolving function of redshift. The solid black line is the full numerical solution obtained by integrating Equation \ref{eq:AngleAveragedFlux} with a Gaussian quadrature technique, and the dashed black line is the same calculation assuming the optically thin $\xibar(z) = \text{constant} = 1$ limit as opposed to $\xibar(z) = \text{constant} = 0$. In order to prevent errors in $J_{\nu}$ at all energies $h\nu \geq 0.2$ keV, the redshift dimensions of $\taubar$ must be sampled at better than $\Delta z = 0.05$ resolution.}
\label{fig:flux_linear_dz}
\end{figure}

The consequences of under-sampling the optical depth are shown in Figure \ref{fig:flux_linear_dz}, which shows the X-ray background spectrum at $z=20$ for a population of $10 \ \Msun$ BHs with multi-color disk (MCD) spectra \citep{Mitsuda1984} and our default set of parameters, which will be described in more detail in Section \ref{sec:Results} (summarized in Table \ref{tab:const_pars}). Soft X-rays are absorbed over small redshift intervals -- in some cases over intervals smaller than those sampled in the optical depth table -- which leads to overestimates of the soft X-ray background intensity. Overestimating the soft X-ray background intensity can lead to significant errors in the resulting heating since soft X-rays are most readily absorbed by the IGM (recall $\sigma_{\nu} \propto \nu^{-3}$ approximately). For a redshift grid with points linearly spaced by an amount $\Delta z = \{0.4, 0.2, 0.1, 0.05 \}$, the errors in $J_{\nu}$ as shown in Figure \ref{fig:flux_linear_dz} correspond to relative errors in the heating rate density, $\eX$, of $\{1.1, 0.44, 0.15, 0.04 \}$. Errors in $\eX$ due to frequency sampling (128 used points here) are negligible (relative error $< 10^{-4}$).

To prevent the errors in $\eX$ associated with under-sampling $\taubar$, we must understand how far X-rays of various energies travel before being absorbed. We estimate a characteristic differential redshift element over which photons are absorbed by assuming a fully neutral medium, and approximate bound-free photo-ionization cross-sections ($\sigma \propto \nu^{-3}$), in which case the optical depth (Eq. \ref{eq:OpticalDepth}) can be written analytically as
\begin{equation}
    \overline{\tau}_{\nu}(z, z^{\prime}) \simeq \left(\frac{\mu}{\nu}\right)^3(1 + z)^{3/2} \left[1 - \left(\frac{1 + z}{1 + z^{\prime}}\right)^{3/2} \right] \label{eq:TauAnalytic} ,
\end{equation} 
where
\begin{equation}
    \mu^3 \equiv \frac{2}{3} \frac{\nHbar \sigma_0 c}{H_0 \sqrt{\Omega_{m,0}}} \left(\nu_{\HI}^3 + y \nu_{\HeI}^3\right) \label{eq:DefinitionMu}.
\end{equation}
Here, $\sigma_0$ is the cross-section at the hydrogen ionization threshold, $h\nu_{\HI}$ and $h\nu_{\HeI}$ are the ionization threshold energies for hydrogen and helium, respectively, $y$ is the primordial helium abundance by number, $H_0$ the Hubble parameter today, and $\Omega_{m,0}$ the matter density relative to the critical density today. 

The characteristic energy $h\mu \simeq 366.5 \ \mathrm{eV}$ may be more familiar as it relates to the mean-free paths of photons in a uniform medium relative to the Hubble length, which we refer to as ``Hubble photons,'' with energy $h\nuHub$,
\begin{align}
    h\nuHub & \simeq h\mu \left[\frac{3}{2}\right]^{1/3} (1 - x_i)^{1/3} (1 + z)^{1/2} \nonumber \\
    & \simeq 1.5 (1 - x_i)^{1/3} \left(\frac{1 + z}{10}\right)^{1/2} \ \mathrm{keV} . \label{eq:HubblePhoton}
\end{align}    

The characteristic differential redshift element of interest (which we refer to as the ``bound-free horizon,'' and denote $\Delta z_{bf}$) can be derived by setting $\tau_{\nu}(z, z^{\prime}) = 1$ and taking $z^{\prime} = z + \Delta z_{bf}$ in Equation \ref{eq:TauAnalytic}. The result is
\begin{equation}
    \Delta z_{bf} \simeq (1 + z) \left\{\left[1 - \left(\frac{\nu/\mu}{\sqrt{1+z}}\right)^{3} \right]^{-2/3} - 1 \right\} . \label{eq:TauNeutral}
\end{equation}    
That is, a photon with energy $h\nu$ observed at redshift $z$ has experienced an optical depth of 1 since its emission at redshift $z + \Delta z_{bf}$ and energy $h\nu [1 + \Delta z_{bf} / (1 + z)]$. Over the interval $10 \lesssim z \lesssim 40$, this works out to be $0.1 \lesssim \Delta z_{bf} \lesssim 0.2$ assuming a photon with frequency $\nu = \mu$.

In order to accurately compute the flux (and thus heating), one must resolve this interval with at least a few points, which explains the convergence in Figure \ref{fig:flux_linear_dz} once $\Delta z \leq 0.1$ for $h\nu \lesssim 350$ eV. We discretize logarithmically in redshift (for reasons that will become clear momentarily) following the procedure outlined in Appendix C of \cite{Haardt1996}, first defining 
\begin{equation}
    x \equiv 1 + z ,
\end{equation}
which allows us to set up a logarithmic grid in $x$-space such that
\begin{equation}
    R \equiv \frac{x_{l+1}}{x_l} = \mathrm{constant}
\end{equation}
where $l = 0,1,2,...n_z-1$. The corresponding grid in photon energy space is
\begin{equation}
    h\nu_n = h\nu_{\min} R^{n-1} ,
\end{equation}
where $h\nu_{\min}$ is the minimum photon energy we consider, and $n=1, 2, ...n_{\nu}$. The number of frequency bins, $n_{\nu}$, can be determined iteratively in order to guarantee coverage out to some maximum emission energy, $h\nu_{\max}$.

The emission frequency, $\nu_{n^{\prime}}$ of a photon observed at frequency $h\nu_n$ and redshift $z_l$, emitted at redshift $z_m$ is then (i.e. a discretized form of Eq. \ref{eq:EmissionFrequency})
\begin{equation}
    \nu_{n^{\prime}} = \nu_n \left(\frac{1+z_m}{1+z_l} \right) \label{eq:EmissionFrequencyDiscrete}
\end{equation}
meaning $\nu_{n^{\prime}}$ can be found in our frequency grid at index
$n^{\prime} = n+m-l$.

The advantage of this approach still may not be immediately obvious, but consider breaking the integral of Equation \ref{eq:AngleAveragedFlux} into two pieces, an integral from $z_l$ to $z_{l+1}$, and an integral from $z_{l+1}$ to $z_{n_z-1}$. In this case, Equation \ref{eq:AngleAveragedFlux} simplifies to
\begin{multline}
    \widehat{J}_{\nu_n} (z_l) = \frac{c}{4\pi} (1 + z_l)^2 \int_{z_l}^{z_{l+1}} \frac{\hat{\upepsilon}_{\nu_{n^{\prime}}}(z^{\prime})}{H(z^{\prime})} e^{-\overline{\tau}_{\nu_n}(z_l, z^{\prime})} dz^{\prime} \\
     + \left(\frac{1+z_l}{1+z_{l+1}}\right)^2 \widehat{J}_{\nu_{n+1}}(z_{l+1}) e^{-\overline{\tau}_{\nu_n}(z_l, z_{l+1})} \label{eq:FluxDiscrete}.
\end{multline}
The first term accounts for ``new'' flux due to the integrated emission of
sources at $z_l \leq z \leq z_{l+1}$, while the second term is the flux due to
emission from all $z > z_{l+1}$, i.e., the background intensity at $z_{l+1}$
corrected for geometrical dilution and attenuation between $z_l$ and
$z_{l+1}$.

Equation \ref{eq:FluxDiscrete} tells us that by discretizing logarithmically
in redshift and iterating from high redshift to low redshift we can keep a
``running total'' on the background intensity. In fact, we must never
explicitly consider the case of $m \neq l+1$, meaning Equation
\ref{eq:EmissionFrequencyDiscrete} is simply $\nu_{n^{\prime}} = R \nu_n =
\nu_{n+1}$. The computational cost of this algorithm is independent of
redshift, since the flux at $z_l$ only ever depends on quantities at $z_l$ and
$z_{l+1}$. Such is not the case for a brute-force integration of Equation
\ref{eq:AngleAveragedFlux}, in which case the redshift interval increases with
time. The logarithmic approach also limits memory consumption, since we need
not tabulate the flux or optical depth in 3-D --- we only ever need to know
the optical depth between redshifts $z_l$ and $z_{l+1}$ --- in addition to the
fact that we can discard the flux at $z_{l+2}$, $J_{\nu}(z_{l+2})$, once we
reach $z_l$. A linear discretization scheme would require 3-D optical depth
tables with $n_{\nu} n_z^2$ elements, which translates to tens of Gigabytes of
memory for the requisite redshift resolution (to be discussed in the next
subsection).

Finally, linear discretization schemes prevent one from keeping a ``running
total'' on the background intensity, since the observed flux at redshift $z_l$
and frequency $\nu_n$ cannot (in general) be traced back to rest frame
emission from redshifts $z_{l^{\prime}}$ or frequencies $\nu_{n^{\prime}}$
within the original redshift and frequency grids (over $l$ and $n$). The
computational cost of performing the integral in Equation
\ref{eq:AngleAveragedFlux} over all redshifts $z^{\prime} > z$ is prohibitive,
as noted by previous authors \citep[e.g.,][]{Mesinger2011}.

\subsection{Accuracy \& Expense} \label{sec:Performance}
\begin{figure}
\includegraphics[width=0.49\textwidth]{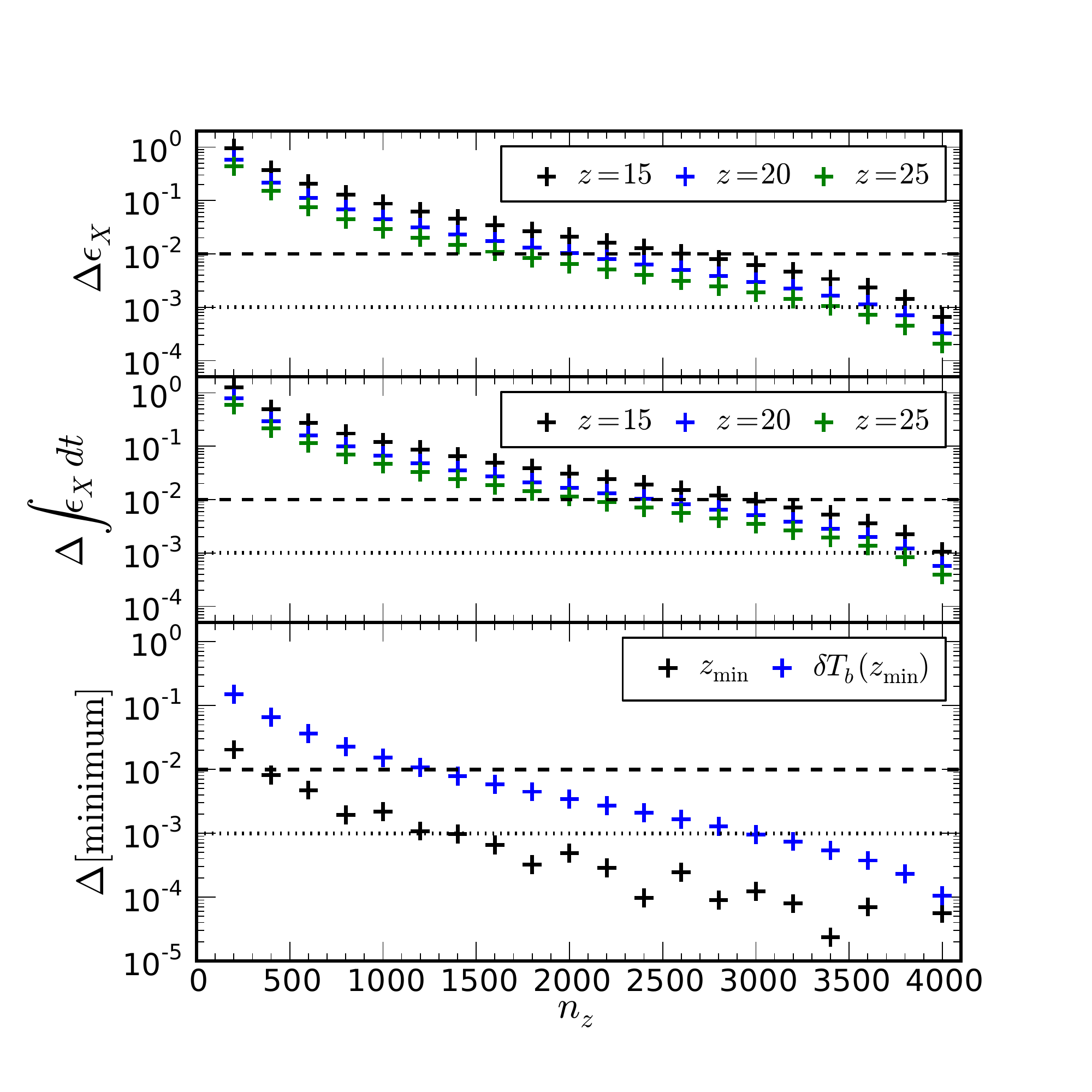}
\caption{Accuracy of presented algorithm. \textit{Top}: Relative error in the heating rate density, $\eX$, as a function of the number of redshift points, $n_z$, used to sample $\taubar$, as compared to a brute-force solution to Equation \ref{eq:HeatingRateDensity} using a double Gaussian quadrature integration scheme. \textit{Middle}: Relative error in the cumulative heating as a function of $n_z$. \textit{Bottom}: Relative error in the position of the 21-cm minimum, in redshift (black crosses) and amplitude (blue crosses). Dotted and dashed lines indicate 0.1\% and 1\% errors, respectively.}
\label{fig:performance}
\end{figure}

The accuracy of this approach is shown in Figure \ref{fig:performance} as a
function of the number of redshift bins in the optical depth lookup table,
$n_z$. Errors in the heating rate density (top), and cumulative heating
(middle), $\Delta \int \eX dt$, drop below 0.1\% at all $10 \leq z \leq 40$
once $n_z \gtrsim 4000$, at which time errors in the position of the 21-cm
minimum (bottom) are accurate to $\sim 0.01$\%. Given this result, all
calculations reported in Section \ref{sec:Results} take $n_z=4000$. For
reference, errors of order 0.1\% correspond to $\sim 0.1$ mK errors in the
amplitude of the 21-cm minimum in our reference model, which we will soon find
is much smaller than the changes induced by physical effects.

\begin{figure*}
\includegraphics[width=0.98\textwidth]{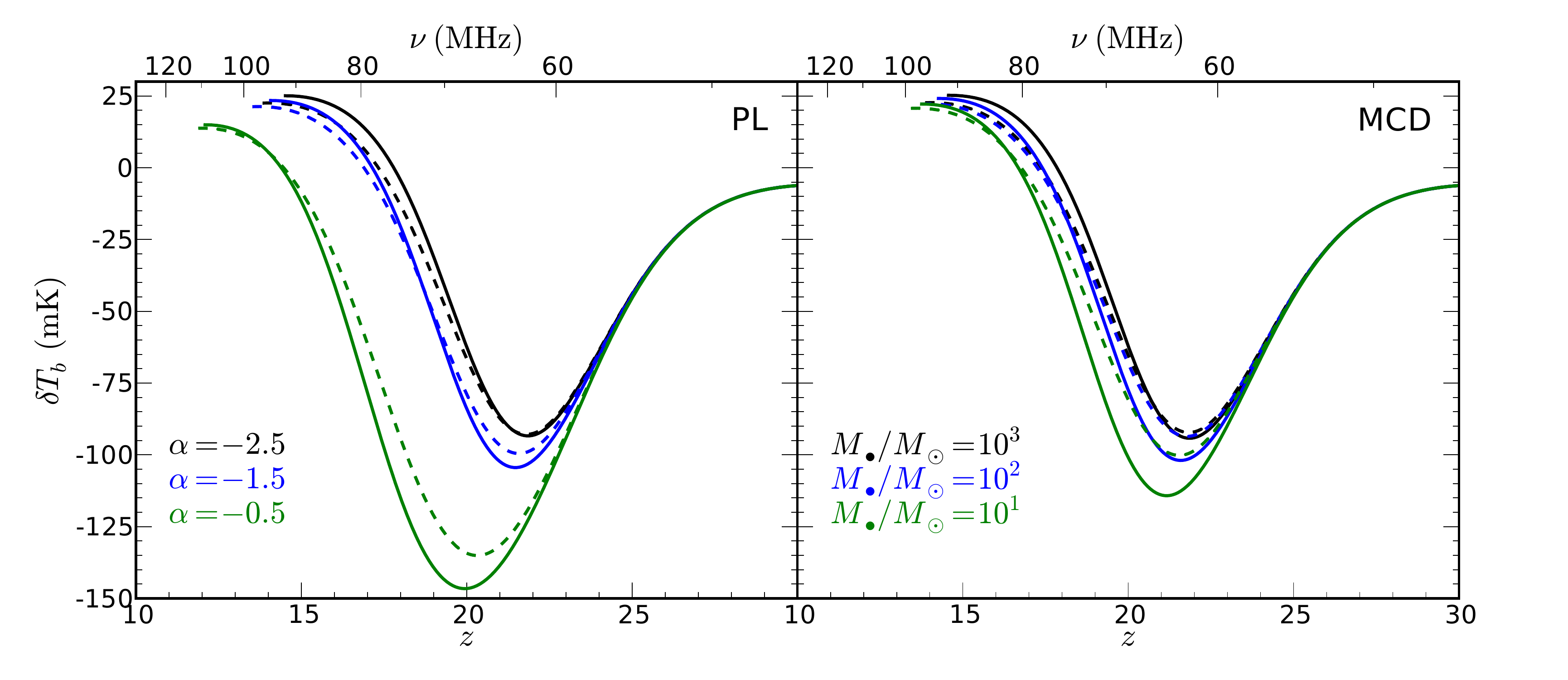}
\caption{Testing the approximation of Equations \ref{eq:xiX} and \ref{eq:xiX_heat}. Dashed lines represent the approximate solutions, while
solid lines represent the full solution for the global 21-cm signal using the
procedure outlined in Section \ref{sec:Methods}. \textit{Left}: X-ray sources
are assumed to have power-law (PL) SEDs with spectral index $\alpha$,
extending from 0.2 to 30 keV. \textit{Right}: X-ray sources are assumed to
have multi-color disk (MCD) SEDs \citep{Mitsuda1984}. All sources have been
normalized to have the same luminosity density above 0.2 keV ($3.4 \times
10^{40} \mathrm{erg} \ \mathrm{s}^{-1} \ (\Msun \ \mathrm{yr}^{-1})^{-1}$),
and all calculations are terminated once the emission peak ($ 12 \lesssim z
\lesssim 14 $) has been reached. For the hardest sources of X-rays considered
(left: $\alpha = -0.5$, right: $\Mbh = 10 \ \Msun$), the global 21-cm minimum
is in error by up to $\sim 15$ mK in amplitude and $\Delta z \simeq 0.5$ in
position when Equation \ref{eq:xiX} is used to compute $\eX$.}
\label{fig:pl_vs_mcd}
\end{figure*}

Many previous studies avoided the expense of Equation
\ref{eq:AngleAveragedFlux} by assuming that a constant fraction of the X-ray
luminosity density is deposited in the IGM as heat
\citep[e.g.,][]{Furlanetto2006}. A physically-motivated approximation is to
assume that photons with short mean free paths (e.g., those that experience
$\overline{\tau}_{\nu} \leq 1$) are absorbed and contribute to heating, and
all others do not \citep[e.g.,][]{Mesinger2011}. This sort of ``step
attenuation'' model was recently found to hold fairly well in the context of a
fluctuating X-ray background, albeit for a single set of model parameters
\citep{Mesinger2009b}.

An analogous estimate for the heating caused by a uniform radiation background
assumes that photons with mean free paths shorter than a Hubble length are absorbed, and all others are not. We define $\xi_X$ as the fraction of the bolometric luminosity density that is absorbed locally, which is given by
\begin{equation}
    \xi_X(z) \approx \int_{\nu_{\min}}^{\nuHub} I_{\nu} d\nu \left(\int_{\nu_{\min}}^{\nu_{\max}} I_{\nu} d\nu \right)^{-1} \label{eq:xiX} ,
\end{equation}
where $h\nuHub$ is given by Equation \ref{eq:HubblePhoton}. There are approximate analytic solutions to the above equation for power-law sources (would be
exact if not for the upper integration limit, $\nuHub$), though $\xi_X$ must be computed numerically for the MCD spectra we consider. We take $h\nu_{\min}
= 200 \ \mathrm{eV}$ and $h\nu_{\max} = 30 \ \mathrm{keV}$ for the duration of
this paper. The heating rate density associated with a population of objects described by $\xi_X$ and $\Lbol$ is 
\begin{equation}
    \eX(z) = \xi_{X}(z) \Lbol(z) \fheat \label{eq:xiX_heat} 
\end{equation}
where $\fheat$ is the fraction of the absorbed energy that is deposited as heat. Because there is no explicit dependence on photon energy in this approximation, we use the fitting formulae of \citet{Shull1985} to compute $\fheat$.

The consequences of using Equations \ref{eq:xiX} and \ref{eq:xiX_heat} for the
global 21-cm signal are illustrated in Figure \ref{fig:pl_vs_mcd}. Steep
power-law sources can be modeled quite well (signal accurate to 1-2 mK) using
Equations \ref{eq:xiX} and \ref{eq:xiX_heat} since a large fraction of the
X-ray emission occurs at low energies. In contrast, heating by sources with
increasingly flat (decreasing spectral index $\alpha$) spectra is poorly
modeled by Equations \ref{eq:xiX} and \ref{eq:xiX_heat}, inducing errors in
the global 21-cm signal of order $\sim 5$ mK ($\alpha=-1.5$) and $\sim 15$ mK
($\alpha=-0.5$). The same trend holds for heating dominated by sources with a
MCD spectrum, in which case harder spectra correspond to less massive BHs. We
will see in the next section that these errors are comparable to the
differences brought about by real changes in the SED of X-ray sources.

\section{ACCRETING BLACK HOLES IN THE EARLY UNIVERSE} \label{sec:Results}
\begin{figure}
\includegraphics[width=0.49\textwidth]{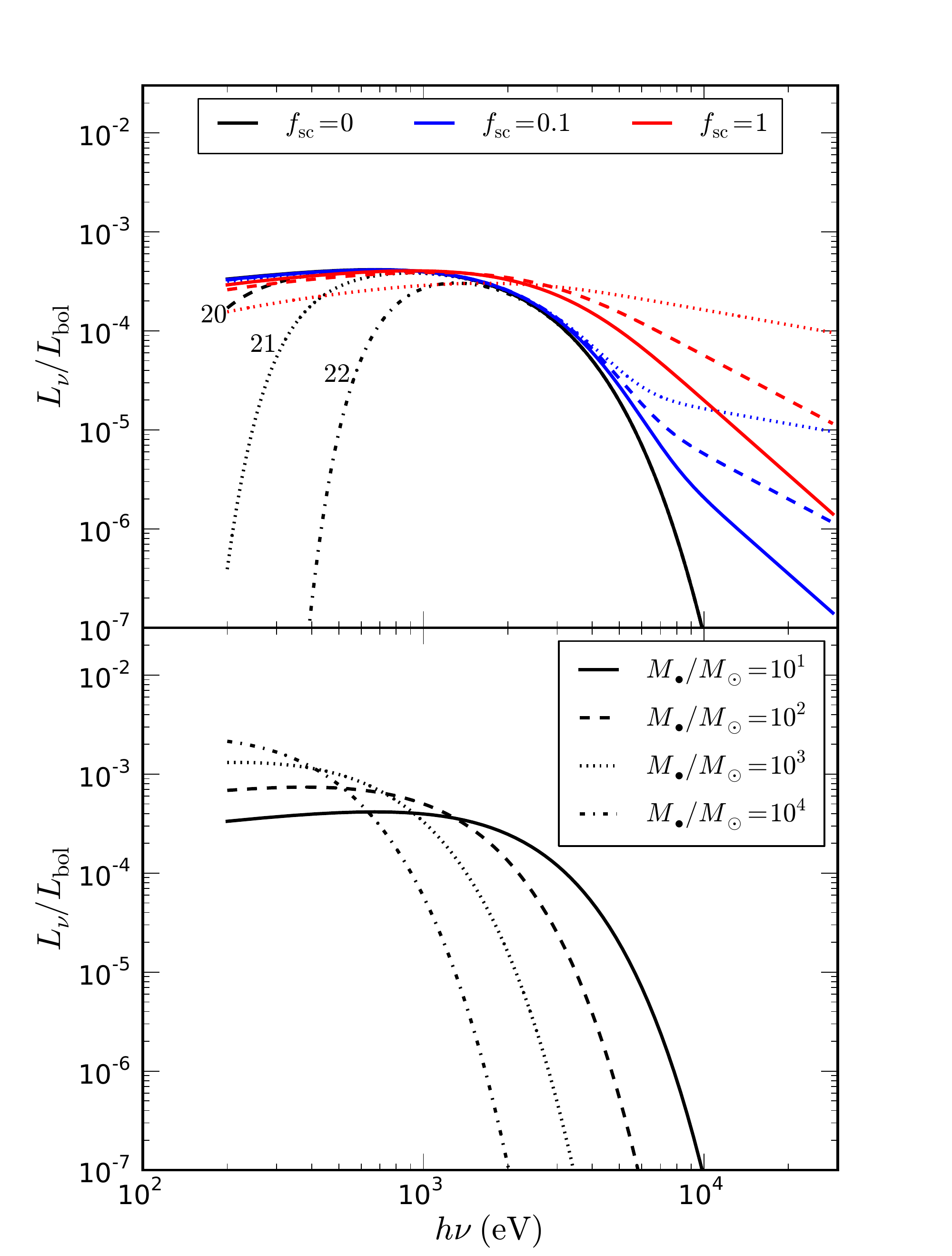}
\caption{Subset of SEDs used in this work. \textit{Top panel:} Assuming $\Mbh = 10 \ \Msun$, varying the fraction of disk photons scattered into the high energy power-law tail, $\fsc$, and the spectral index of the resulting high energy emission, $\alpha$, using the SIMPL model \citep{Steiner2009}. Solid, dashed, dotted, and dash-dotted black lines represent neutral absorption corresponding to $\NHI / \mathrm{cm}^{-2} = 0, 10^{20}, 10^{21}$, and $10^{22}$, respectively. Solid and dashed lines of different colors correspond to high energy emission with power-law indices of $\alpha=-2.5$ and $\alpha=-1.5$, respectively, with the color indicating $\fsc$ as shown in the legend. \textit{Bottom panel:} Pure MCD SEDs for $\Mbh = 10-10^4 \ \Msun$, with no intrinsic absorption or Comptonization of the disk spectrum. The solid black line is our reference model, and is the same in both panels.}
\label{fig:seds}
\end{figure}

Using the algorithm presented in the previous section, we now investigate the effects of varying four parameters that govern the SED of an accreting BH: (1) the mass of the BH, $\Mbh$, which determines the characteristic temperature of an optically thick geometrically thin disk \citep{ShakuraSunyaev1973}, (2) the fraction of disk photons that are up-scattered \citep{Shapiro1976} by a hot electron corona, $\fsc$, (3) the power-law index\footnote{We define the spectral index as $L_{\nu} \propto \nu^{\alpha}$, where $L_{\nu}$ is a specific luminosity proportional to the \textit{energy} of a photon with frequency $\nu$, per logarithmic frequency interval $d\nu$.} of the resulting emission, $\alpha$, which describes respectively \citep[using the SIMPL model;][]{Steiner2009}, and (4) the column density of neutral hydrogen that lies between the accreting system and the IGM, $\NHI$. Because we assume $\xHII = \xHeII$, the absorbing column density actually has an optical depth of $\tau_{\nu} = \NHI \sigma_{\nu, \HI} (1 + y \sigma_{\nu, \HeI} / \sigma_{\nu, \HI})$, where $y$ is the primordial helium abundance by number, and $\sigma_{\nu}$ is the bound-free absorption cross section for $\HI$ and $\HeI$. A subset of the spectral models we consider are shown in Figure \ref{fig:seds}. Note that more efficient Comptonization (i.e., increasing $\fsc$) and strong neutral absorption (increased $\NHI$) act to harden the intrinsic disk spectrum (top panel), while increasing the characteristic mass of accreting BHs acts to soften the spectrum (bottom panel).

%

To compute the X-ray heating as a function of redshift, $\eX(z)$, we scale our SED of choice to a co-moving (bolometric) luminosity density assuming that a constant fraction of gas collapsing onto halos accretes onto BHs, i.e.,
\begin{equation}
    \rhobhdot(z) = \fbh \rhobbar \frac{d \fcoll(\Tmin)}{dt} . \label{eq:rhobhdot}
\end{equation}
Assuming Eddington-limited accretion, we obtain a co-moving bolometric ``accretion luminosity density,''
\begin{align}
    \mathcal{L}_{\acc} = 6.3 & \times 10^{40} \ \mathrm{erg} \ \mathrm{s}^{-1} \ \mathrm{cMpc}^{-3} \nonumber \\
    & \times \left(\frac{0.9}{\xi_{\mathrm{acc}}} \right) \left(\frac{\rhobhdot(z)}{10^{-6} \ M_{\odot} \ \mathrm{yr}^{-1} \ \mathrm{cMpc}^{-3}} \right)  \label{eq:Lacc} ,
\end{align}
where
\begin{equation}
    \xi_{\mathrm{acc}} = \frac{1 - \eta}{\eta} \fedd
\end{equation}
and $\eta$ and $\fedd$ are the radiative efficiency and Eddington ratio, respectively. To be precise, $\fedd$ represents the product of the Eddington ratio and duty cycle, i.e., what fraction of the time X-ray sources are actively accreting, which are completely degenerate. This parameterization is very similar to that of \citet{Mirabel2011}, though we do not explicitly treat the binary fraction, and our expression refers to the bolometric luminosity density rather than the 2-10 keV luminosity density. Our model for the co-moving X-ray emissivity is then
\begin{equation}
    \enu(z) = \mathcal{L}_{\acc}(z) \frac{I_{\nu}}{h\nu} , \label{eq:Emissivity}
\end{equation}
where $I_{\nu}$ once again represents the SED of X-ray sources, and is normalized such that $\int_0^{\infty} I_{\nu} d\nu = 1$. Power-law sources must truncate the integration limits in this normalization integral so as to avoid divergence at low energies, though MCD models do not, since the soft X-ray portion of the spectrum is limited by the finite size of the accretion disk (which we take to be $r_{\max} = 10^3 \ R_g$, where $R_g = G\Mbh/c^2$).

It is common in the 21-cm literature to instead relate the co-moving X-ray luminosity density, $L_X$, to the star formation rate density, $\rhostardot$, as 
\begin{equation}
    L_X = c_X f_X \rhostardot(z) ,
\end{equation}
where the normalization factor $c_X$ is constrained by observations of nearby star forming galaxies \citep[e.g.,][]{Grimm2003, Ranalli2003, Gilfanov2004}, and $f_X$ parameterizes our uncertainty in how the $L_X-\text{SFR}$ relation evolves with redshift. The detection of a 21-cm signal consistent with $f_X > 1$ could provide indirect evidence of a top-heavy stellar initial mass function (IMF) at high-$z$ since $f_X$ encodes information about the abundance of high-mass stars and the binary fraction, both of which are expected to increase with decreasing metallicity. 

\begin{figure*}
\includegraphics[width=0.98\textwidth]{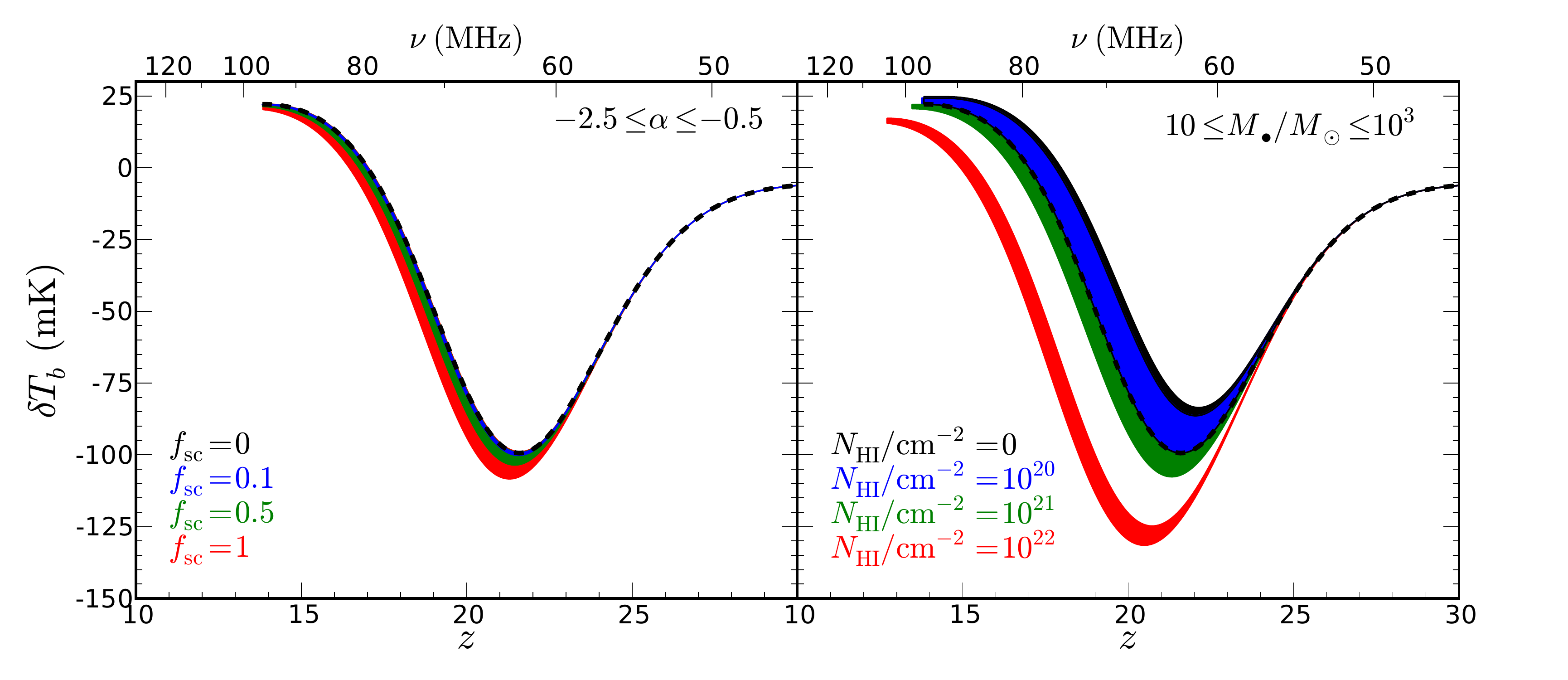}
\caption{Evolution of the 21-cm brightness temperature for different BH SED
models. \textit{Left:} Effects of coronal physics, parameterized by the
fraction of disk photons up-scattered by a hot electron corona, $\fsc$, and
the resulting spectral index of up-scattered emission, $\alpha$, using the
SIMPL Comptonization model of \citet{Steiner2009}. The colors correspond to
different values of $\fsc$, while the width of each band represents models
with $-2.5 \leq \alpha \leq -0.5$ (the upper edge of each band corresponds to
the softest SED at fixed $\fsc$, in this case $\alpha=-2.5$). \textit{Right:}
Effects of BH mass and neutral absorbing column. Colors correspond to $\NHI$,
while the width of each band represents models with $10 \leq \Mbh / \Msun \leq
10^3$ (the upper edge of each band corresponds to the softest SED at fixed
$\NHI$, in this case $\Mbh=10^3 \ \Msun$). The dashed black line is our
reference ``pure MCD'' model with $\Mbh = 10 \ \Msun$. The black and blue
regions overlap considerably, indicating that absorbing columns of $\NHI
\gtrsim 10^{20} \ \mathrm{cm}^2$ are required to harden the spectrum enough to
modify the thermal history. Every realization of the signal here has the exact
same ionization history, $\Lya$ background history, and BH accretion history.
As in Figure \ref{fig:pl_vs_mcd}, all calculations are terminated once the
peak in emission is reached. Coronal physics influences the global 21-cm
minimum at the $\lesssim 10$ mK level, while $\Mbh$ is a 10-20 mK effect and
$\NHI$ is potentially a $\sim 50$ mK effect.}
\label{fig:simpl_21cm}
\end{figure*}

However, assumptions about the SED of X-ray sources are built-in to the
definition of $f_X$. The standard value of $c_X = 3.4 \times 10^{40}
\mathrm{erg} \ \mathrm{s}^{-1} \ (\Msun \ \mathrm{yr}^{-1})^{-1}$
\citep{Furlanetto2006} is an extrapolation of the $2-10$ keV $L_X-\text{SFR}$
relation of \citet{Grimm2003}, who found $L_{2-10 \ \mathrm{keV}} = 6.7 \times
10^{39} \mathrm{erg} \ \mathrm{s}^{-1} \ (\Msun \ \mathrm{yr}^{-1})^{-1}$, to
all energies $h\nu > 200 \ \mathrm{eV}$ assuming an $\alpha=-1.5$ power-law
spectrum. This means any inferences about the stellar IMF at high-$z$ drawn
from constraints on $f_X$ implicitly assume an $\alpha=-1.5$ power-law
spectrum at photon energies above 0.2 keV. Because our primary interest is in
SED effects, we avoid the $f_X$ parameterization and keep the normalization of
the X-ray background (given by $\rhobhdot / \xi_{\mathrm{acc}}$) and its SED
($I_{\nu}$) separate. We note that if one adopts a pure MCD spectrum (i.e.,
$\fsc = \NHI = 0$) for a $10 \ \Msun$ BH and set $\fbh=10^{-5}$ (as
in our reference model), the normalization of Equation \ref{eq:Lacc}
corresponds to $\fX \approx 2 \times 10^3$ assuming $c_X = 2.61 \times 10^{39}
\cXunits$ \citep{Mineo2012}. Despite this enhancement in the total X-ray
luminosity density, our reference model produces an absorption trough at $z
\approx 22$ and $\delta T_b \approx -100$ mK, similar to past work that
assumed $\fX = 1$. This is a result of our choice for the reference spectrum,
a multi-color disk, which is much harder than the $\alpha=-1.5$ power law
spectrum originally used to define $f_X$.

Our main result is shown in Figure \ref{fig:simpl_21cm}. The effects of the
coronal physics parameters $\fsc$ and $\alpha$ are shown in the left panel,
and only cause deviations from the reference model if $\fsc > 0.1$ (for any
$-2.5 \leq \alpha \leq -0.5$). Increasing $\fsc$ and decreasing $\alpha$ act
to harden the spectrum, leading to a delay in the onset of heating and thus
deeper absorption feature. With a maximal value of $\fsc = 1$ and hardest
power-law SED of $\alpha=-0.5$, the absorption trough becomes deeper by $\sim
10 \ \mathrm{mK}$. In the right panel, we adopt $\fsc = 0.1$ and
$\alpha=-1.5$, and turn our attention to the characteristic mass of accreting
BHs and the neutral absorbing column, varying each by a factor of 100, each of
which has a more substantial impact individually on the 21-cm signal than
$\fsc$ and $\alpha$. The absorption trough varies in amplitude by up to $\sim
50$ mK and in position by $\Delta z \approx 2$ from the hardest SED ($\Mbh =
10 \ \Msun$, $\NHI = 10^{22} \ \mathrm{cm}^{-2}$) to softest SED ($\Mbh = 10^3
\ \Msun$, $\NHI = 0 \ \mathrm{cm}^{-2}$) we consider. The absorbing column
only becomes important once $\NHI \gtrsim 10^{20} \ \mathrm{cm}^{-2}$.

Our study is by no means exhaustive. Table \ref{tab:const_pars} lists
parameters held constant for the calculations shown in Figure
\ref{fig:simpl_21cm}. Our choices for several parameters in Table
\ref{tab:const_pars} that directly influence the thermal history will be
discussed in the next section. While several other parameters could be
important in determining the locations of 21-cm features, for instance,
$\Nion$ is likely $\gg 4000$ for Population III (PopIII) stars
\citep[e.g.,][]{Bromm2001,Schaerer2002,Tumlinson2003}, we defer a more
complete exploration of parameter space, and assessment of degeneracies
between parameters, to future work.

\begin{table}\scriptsize
\begin{tabular}{ | l | l | l | }
\hline
Parameter & Value & Description \\
\hline    
\texttt{hmf} &  PS  & Halo mass function \\
$T_{\min}$ &$10^4 \ \mathrm{K}$ & Min. virial temperature of star-forming haloes \\ 
$\mu$ & 0.61 & Mean molecular weight of collapsing gas \\
$\fstar$ & $10^{-1}$ & Star formation efficiency \\
$\fbh$ & $10^{-5}$ & Fraction of collapsing gas accreted onto BHs \\
$N_{\mathrm{LW}}$  & 9690 & Photons per stellar baryon with $\nu_{\alpha} \leq \nu \leq \nu_{\mathrm{LL}}$ \\
$N_{\mathrm{ion}}$  & 4000 & Ionizing photons emitted per stellar baryon \\
$f_{\mathrm{esc}}$  & $0.1$ & Escape fraction \\
$r_{\mathrm{in}}$ & $6 \ R_g$ & Radius of inner edge of accretion disk \\
$r_{\max}$ & $10^3 \ R_g$ & Max. radius of accretion disk \\
$\eta$ & $0.1$ & Radiative efficiency of accretion \\
$\fedd$ & $0.1$ & Product of Eddington ratio and duty cycle \\
$h\nu_{\min}$ & $0.2$ keV & Softest photon considered \\
$h\nu_{\max}$ & $30$ keV & Hardest photon considered \\
\hline
\end{tabular}
\caption{Parameters held constant in this work. Note that PS in the first row refers to the original analytic halo mass function derived by \citet{PressSchecter1974}. Our reference model adopts this set of parameters and a pure MCD spectrum (i.e., $\NHI = \fsc = 0$) assuming a characteristic BH mass of $\Mbh = 10 \ \Msun$.}
\label{tab:const_pars}
\end{table}

\section{DISCUSSION} \label{sec:Discussion}
The findings of the previous section indicate that uncertainty in the SED of X-ray sources at high-$z$ could be a significant complication in the interpretation of upcoming 21-cm measurements. Details of Comptonization are a secondary effect in this study, though still at the level of measurement errors predicted by current signal extraction algorithms \citep[likely $\sim 10$ mK for the absorption trough;][]{Harker2012}. The characteristic mass of accreting BHs, $\Mbh$, and the amount of absorption intrinsic to BH host galaxies, parameterized by a neutral hydrogen column density $\NHI$, influence the signal even more considerably. In this section, we examine these findings in the context of other recent studies and discuss how our methods and various assumptions could further influence our results.

\subsection{An Evolving IGM Optical Depth}
Central to our approach to solving Equation \ref{eq:AngleAveragedFlux} is the ability to tabulate the IGM optical depth (Eq. \ref{eq:OpticalDepth}). This requires that we assume a model for the ionization history \textit{a-priori}, even though the details of the X-ray background will in general influence the ionization history to some degree\footnote{Evolution of the volume filling factor of HII regions, $x_i$, is the same in each model we consider because we have not varied the number of ionizing photons emitted per baryon of star formation, $\Nion$, or the star formation history, parameterized by the minimum virial temperature of star-forming haloes, $\Tmin$, and the star formation efficiency, $\fstar$. X-rays are only allowed to ionize the bulk IGM in our formalism, whose ionized fraction is $x_e \lesssim 0.1$\% at all $z \gtrsim 12$ in our models, meaning $\xibar \approx x_i$. The midpoint of reionization occurs at $z \simeq 10.8$ in each model we consider.}. Because we focus primarily on 21-cm features expected to occur at $z > 10$, we assume $\overline{x}_i = x_e = 0$ at all $z > 10$ when generating $\taubar(z, z^{\prime})$.

\begin{figure}
\includegraphics[width=0.49\textwidth]{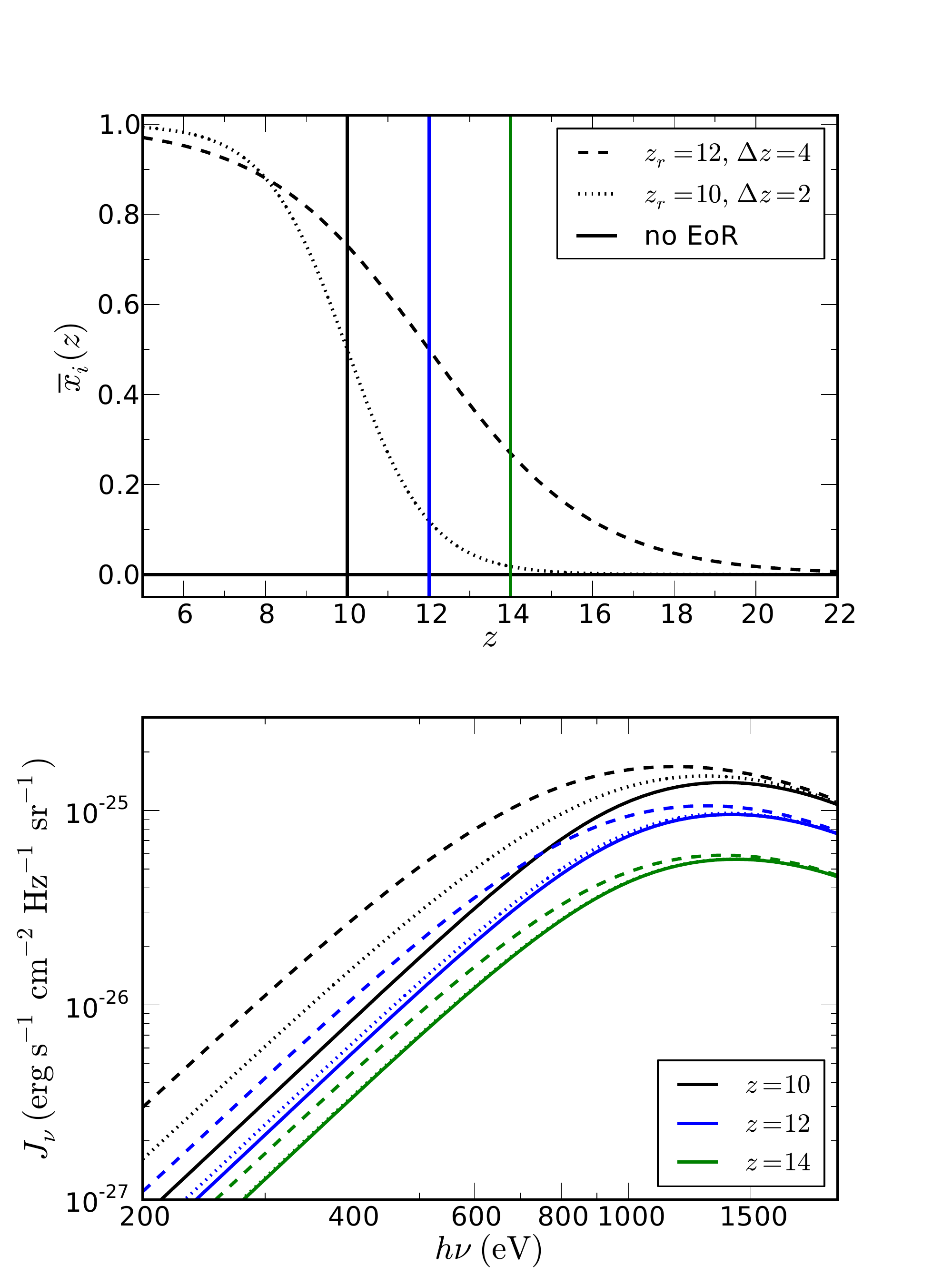}
\caption{Consequences of the $\xibar = \text{constant} = 0$ approximation on the background radiation field for our reference model (see Table \ref{tab:const_pars}). \textit{Top}: \textit{tanh} ionization histories considered, i.e., $\xibar(z) \propto \text{tanh}((z - z_r) / \Delta z)$. \textit{Bottom}: Angle-averaged background intensity, $J_{\nu}$, at $z=10, 12$ and $14$ (black, blue, green) assuming a neutral IGM for all $z$ (solid), compared to increasingly early and extended reionization scenarios (dotted and dashed). Errors in the background intensity due to the $\xibar = \text{constant} = 0$ could be important at $z \lesssim 14$, assuming early and extended reionization scenarios (e.g., $z_r = 12$, $\Delta z = 4$), though by this time the global 21-cm signal is likely insensitive to the thermal history.}
\label{fig:evolving_xi}
\end{figure}

The effects of this approximation are shown in Figure \ref{fig:evolving_xi}, in which we examine how different ionization histories (and thus IGM opacities) affect the background flux, $J_{\nu}$. Because we assume a neutral IGM for all $z \geq 10$, we always underestimate the background flux, since an evolving IGM optical depth due to reionization of the IGM allows X-rays to travel further than they would in a neutral medium. The worst-case-scenario for this $\xibar(z) = 0$ approximation occurs for very extended ionization histories (blue line in top panel of Figure \ref{fig:evolving_xi}), in which case the heating rate density at $z = \{10, 12, 14\}$ is in error by factors of $\{1.2, 0.5, 0.2\}$. Because the 21-cm signal is likely insensitive to $\eX$ once reionization begins\footnote{Though ``cold reionization'' scenarios have not been completely ruled out, recent work is inconsistent with a completely unheated $z \approx 8$ IGM \citep{Parsons2014}.}, we suspect this error is negligible in practice. As pointed out in \citet{Mirocha2013}, the 21-cm emission feature can serve as a probe of $\eX$ so long as independent constraints on the ionization history are in hand. In this case, we would simply tabulate $\overline{\tau}_{\nu}$ using the observational constraints on $\xibar(z)$, and mitigate the errors shown in Figure \ref{fig:evolving_xi}. Our code could also be modified to compute the optical depth on-the-fly once $\xibar$ exceeds a few percent, indicating the beginning of the EoR.

\subsection{Neutral Absorption}
Our choice of $\NHI$ is consistent with the range of values adopted in the
literature in recent years \citep[e.g.,][]{Mesinger2013a}, which are chosen to
match constraints on neutral hydrogen absorption seen in high-$z$ gamma ray
burst spectra \citep[which can also be explained if reionization is patchy or
not complete by $z \approx 7$;][]{Totani2006, Greiner2009}. If we assume that
the absorbing column is due to the host galaxy ISM, then it cannot be used
solely to harden the X-ray spectrum -- it must also attenuate soft UV photons
from stars, and thus be related to the escape fraction of ionizing radiation,
$\fesc$. In the most optimistic case of a PopIII galaxy (which we take
to be a perfect blackbody of $10^5$ K), an absorbing column of $\NHI =
10^{18.5} \ \mathrm{cm}^{-2}$ corresponds to $\fesc \simeq 0.01$, meaning
every non-zero column density we investigated in Figure \ref{fig:simpl_21cm}
would lead to the attenuation of more than 99\% of ionizing stellar radiation,
thus inhibiting the progression of cosmic reionization considerably.

An alternative is to assume that the absorbing column is intrinsic to
accreting systems, though work on galactic X-ray binaries casts doubt on such
an assumption. \citet{Miller2009} monitored a series of photoelectric
absorption edges during BH spectral state transitions, and found that while
the soft X-ray spectrum varied considerably, the column densities inferred by
the absorption edges remained roughly constant. This supports the idea that
evolution in the soft X-ray spectrum of X-ray binaries arises due to evolution
in the source spectrum, and that neutral absorption is dominated by the host
galaxy ISM.

For large values of $\NHI$, reionization could still proceed if the
distribution of neutral gas in (at least some) galaxies were highly
anisotropic. Recent simulations by \citet{Gnedin2008} lend credence to this
idea, displaying order-of-magnitude deviations in the escape fraction
depending on the propagation direction of ionizing photons -- with radiation
escaping through the polar regions of disk galaxies preferentially.
\citet{Wise2009} performed a rigorous study of ionizing photon escape using
simulations of both idealized and cosmological haloes, reaching similar
conclusions extending to lower halo masses. The higher mass halos in the
\citet{Wise2009} simulation suite exhibited larger covering fractions of high
column density gas (e.g., Figure 10), which could act to harden the spectrum
of such galaxies, in addition to causing very anisotropic HII regions.

If there existed a population of miniquasars powered by intermediate mass BHs,
and more massive BHs at high-$z$ occupy more massive haloes, then more massive
haloes should have softer X-ray spectra (see Figure \ref{fig:seds}) and thus
heat the IGM more efficiently. However, if they also exhibit larger covering
fractions of high column density gas, the soft X-ray spectrum will be
attenuated to some degree -- perhaps enough to mimic an intrinsically harder
source of X-rays. This effect may be reduced in galaxies hosting an AGN, since
X-rays partially ionize galactic gas and thus act to enhance the escape
fraction of hydrogen- and helium- ionizing radiation \citep{Benson2013}.
Ultimately the 21-cm signal only probes the volume-averaged emissivity, so if
soft X-ray sources reside in high-mass haloes, they would have to be very
bright to compensate for their rarity, and to contribute substantially to the
heating of the IGM.

Lastly, it is worth mentioning that the hardness of the radiation field
entering the ``neutral'' bulk IGM is not the same as that of the radiation
field leaving the galaxy (whose edge is typically defined as its virial
radius) since our model treats HII regions and the bulk IGM separately. As a
result, there is an extra step between the intrinsic emission (that leaving
the virial radius) and the IGM: of the photons that escape the virial radius,
what fraction of them (as a function of frequency) contribute to the growth of
the galactic HII region? The IGM penetrating radiation field is hardened as a
result, and could become even harder and more anisotropic based on the
presence or absence of large scale structure such as dense sheets and
filaments\footnote{In fact, the metagalactic background could be even harder
than this, given that soft X-rays are absorbed on small scales and thus may
not deserve to be included in a ``global'' radiation background.
\citet{Madau2004} argued for $E_{\min}=150$ eV since 150 eV photons have a
mean-free path comparable to the mean separation between sources in their
models, which formed in $3.5\sigma$ density peaks at $z \sim 24$. However, for
rare sources, a global radiation background treatment may be insufficient
\citep[e.g.,][]{Davies2013}. We chose $E_{\min} = 0.2$ keV to be consistent
with other recent work on the 21-cm signal \citep[e.g.,][]{Pritchard2012}, but
clearly further study is required to determine reasonable values for this
parameter. At least for large values of $\NHI$, the choice of $E_{\min}$ is
irrelevant.}. Additionally, sources with harder spectra lead to more spatially
extended ionization fronts, whose outskirts could be important sources of
21-cm emission \citep[e.g.,][]{Venkatesan2011}.

\subsection{Accretion Physics}
We have assumed throughout a radiative efficiency of $\eta = 0.1$, which is
near the expected value for a thin disk around a non-spinning BH assuming the
inner edge of the disk corresponds to the innermost stable circular orbit,
i.e, $r_{\text{in}} = r_{\text{isco}} = 6 R_g$. The radiative efficiency is
very sensitive to BH spin, varying between $0.05 \leq \eta \leq 0.4$
\citep{Bardeen1970} from maximal retrograde spin (disk and BH angular momentum
vectors are anti-parallel), to maximal prograde spin (disk and BH ``rotate''
in the same sense). While the spin of stellar mass BHs is expected to be
more-or-less constant after their formation \citep{King1999}, the spin
distribution at high-$z$ is expected to be skewed towards large values of the
spin parameter, leading to enhanced radiative efficiencies $\eta > 0.1$
\citep{Volonteri2005}.

Our choice of $\fedd = 0.1$ is much less physically motivated, being that it
is difficult both to constrain observationally and predict theoretically. For
X-ray binaries, $\fedd$ should in general be considered not just what fraction
of time the BH is actively accreting, but what fraction of the time it is in
the high/soft state when the MCD model is appropriate. We ignore this for now
as it is poorly constrained, but note that the emission during the high/soft
state could dominate the heating even if more time is spent in the low/hard
state simply because it is soft X-rays that dominate the heating.

While we don't explicitly attempt to model nuclear BHs, Equation
\ref{eq:rhobhdot} could be used to model their co-moving emissivity. Note,
however, that this model is not necessarily self-consistent. We have imposed
an accretion history via the parameters $\fbh$ and $\Tmin$, though the
Eddington luminosity density depends on the mass density of BHs. For extreme
models (e.g., large values of $\fbh$), the mass density of BHs required to
sustain a given accretion luminosity density can exceed the mass density
computed via integrating the accretion rate density over time. To render such
scenarios self-consistent, one must require BH formation to cease or the
ejection rate of BHs from galaxies to become significant (assuming ejected BHs
no longer accrete), or both. The value of $\fbh$ we adopt is small enough that
we can neglect these complications for now, and postpone more detailed studies
including nuclear BHs to future work.

\subsection{Choosing Representative Parameter Values}
The results of recent population synthesis studies suggest that X-ray binaries
are likely to be the dominant source of X-rays at high-$z$. \citet{Power2013}
modeled the evolution of a single stellar population that forms in an
instantaneous burst, tracking massive stars evolving off the main sequence,
and ultimately the X-ray binaries that form. Taking Cygnus X-1 as a spectral
template, they compute the ionizing luminosity of the population with time
(assuming a Kroupa intial mass function) and find that high-mass X-ray
binaries dominate the instantaneous ionizing photon luminosity starting 20-30
Myr after the initial burst of star formation depending on the binary survival
fraction. \citet{Fragos2013} performed a similar study, but instead started
from the Millenium II simulation halo catalog and applied population synthesis
models to obtain the evolution of the background X-ray spectrum and
normalization from $z \sim 20$ to present day. They find that X-ray binaries
could potentially dominate the X-ray background over AGN (at least from 2-10
keV) at all redshifts higher than $z \sim 5$.

Though our reference model effectively assumes that HMXBs dominate the X-ray
background at high-$z$, supernovae \citep{Oh2001,Furlanetto2004}, accreting
intermediate mass black holes, whether they be solitary ``miniquasars''
\citep[e.g.,][]{Haiman1998,Wyithe2003,Kuhlen2005} or members of binaries, and
thermal bremsstrahlung radiation from the hot interstellar medium of galaxies
could be important X-ray sources as well \citep{Mineo2012b,Pacucci2014}. In
principle, our approach could couple detailed spectral models, composed of
X-ray emission from a variety of sources, to the properties of the IGM with
time, and investigate how the details of population synthesis models, for
example, manifest themselves in the global 21-cm signal. Such studies would be
particularly powerful if partnered with models of the 21-cm angular power
spectrum, observations of which could help break SED-related degeneracies
\citep{Pritchard2007,Mesinger2013a, Pacucci2014}.

\subsection{Helium Effects}
The $x_{\HI} = x_{\HeI}$ approximation we have made throughout is common in
the literature, and has been validated to some extent by the close match in HI
and HeI global ionization histories computed in \citet{Wyithe2003} and
\citet{Friedrich2012}, for example. However, recent studies of the ionization
profiles around stars and quasars \citep[e.g.][]{Thomas2008,Venkatesan2011}
find that more X-ray luminous galaxies have larger HeII regions than HII
regions. Given that the metagalactic radiation field we consider in this work
is even harder than the quasar-like spectra considered in the aforementioned
studies, the HI and HeI fractions in the bulk IGM may differ even more
substantially than they do in the outskirts of HII/HeII regions near quasars.

We have neglected a self-consistent treatment of helium in this work, though
more detailed calculations including helium could have a substantial impact on
the ionization and thermal history. \citet{Ciardi2012} showed that radiative
transfer simulations including helium, relative to their hydrogen-only
counterparts, displayed a slight delay in the redshift of reionization, since
a small fraction of energetic photons are absorbed by helium instead of
hydrogen. The simulations including helium also exhibited an increase in the
IGM temperature at $z \lesssim 10$ due to helium photo-heating. At $z \gtrsim
10$, the volume-averaged temperature in the hydrogen-only simulations was
actually larger due to the larger volume of ionized gas. It is difficult to
compare such results directly to our own, as our interest lies in the IGM
temperature \textit{outside} of ionized regions. Because of this complication,
we defer a more detailed investigation of helium effects to future work.

\section{CONCLUSIONS}
Our conclusions can be summarized as follows:
\begin{enumerate}
    \item Approximate solutions to the cosmological RTE overestimate the heating rate density in the bulk IGM, leading to artificially shallower absorption features in the global 21-cm signal, perhaps by $\sim 15-20 \ \mathrm{mK}$ if sources with hard spectra dominate the X-ray background (Figure \ref{fig:pl_vs_mcd}).
    \item Brute-force solutions are computationally expensive, which limits parameter space searches considerably. The discretization scheme of \citet{Haardt1996} is fast, though exquisite redshift sampling is required in order to accurately model X-ray heating (Figure \ref{fig:performance}).
    \item More realistic X-ray spectra are harder than often used power-law treatments (Figure \ref{fig:seds}), and thus lead to deeper absorption features in the global 21-cm signal at fixed bolometric luminosity density. While the details of coronal physics can harden a ``pure MCD'' spectrum enough to modify the global 21-cm absorption feature at the $\sim 10 \ \mathrm{mK}$ level (in the extreme case of $\fsc = 1$ and $\alpha=-0.5$), the characteristic mass of accreting BHs (amount of neutral absorption in galaxies) has an even more noticeable impact, shifting the absorption trough in amplitude by $\sim 20 \ (\sim 50) \ \mathrm{mK}$ and in redshift by $\Delta z \approx 0.5 \ (\Delta z \approx 2)$ (Figure \ref{fig:simpl_21cm}). 
    \item Care must be taken when using the local $L_X-\text{SFR}$ relation to draw inferences about the high-$z$ stellar IMF, as assumptions about source SEDs are built-in to the often used normalization factor $f_X$. Even if the high-$z$ X-ray background is dominated by X-ray binaries, the parameters governing how significantly the intrinsic disk emission is processed influence the signal enormously, and could vary significantly from galaxy to galaxy.
\end{enumerate}

Though our code was developed to study the global 21-cm signal, it can be used
as a stand-alone radiation background calculator, whose output could be easily
integrated into cosmological simulation codes to investigate large scale
feedback. It is publicly
available\footnote{\href{https://bitbucket.org/mirochaj/glorb}{https://bitbucket.org/mirochaj/glorb}}, and remains under
active development.

J.M. would like to thank Greg Salvesen, Jack Burns, Steven Furlanetto, Andrei
Mesinger, and John Wise for thoughtful discussions and comments on an earlier
version of this manuscript, Stephen Murray for developing the
\texttt{hmf-calc} code \citep{Murray2013} and being so responsive to questions
regarding its use, and the anonymous referee for providing a thorough review
that helped improve the quality of this paper. J.M. acknowledges partial
funding from The LUNAR consortium (http://lunar.colorado.edu), headquartered
at the University of Colorado, which is funded by the NASA Lunar Science
Institute (via Cooperative Agreement NNA09DB30A) to investigate concepts for
astrophysical observatories on the Moon. This work used the \texttt{JANUS}
supercomputer, which is supported by the National Science Foundation (award
number CNS-0821794) and the University of Colorado Boulder. The \texttt{JANUS}
supercomputer is a joint effort of the University of Colorado Boulder, the
University of Colorado Denver, and the National Center for Atmospheric
Research. This work also made use of Python, including the packages
\texttt{numpy}, \texttt{matplotlib}, \texttt{h5py}, and \texttt{scipy}.

\bibliography{references}
\bibliographystyle{mn2e_short}

\appendix

\section{Analytic Test Problem}
In this section, we test our code with a double power-law form for the X-ray emissivity, $\enu(z) \propto (1 + z)^{\beta} \nu^{\alpha - 1}$, noted by \citet{Meiksin2003} to yield analytic solutions in two important limiting cases. In the optically-thin limit \citep[e.g., the cosmologically-limited (CL) case of][in which $\xibar=1$ at all redshifts]{Meiksin2003}, we find
\begin{multline}
    \widehat{J}_{\nu,\mathrm{CL}}(z) = \frac{c}{4\pi} \frac{\hat{\upepsilon}_{\nu}(z)}{H(z)} \frac{(1 + z)^{9/2-(\alpha + \beta)}}{\alpha+\beta-3/2} \\
    \times \left[(1 + z_f)^{\alpha+\beta-3/2} - (1 + z)^{\alpha+\beta-3/2}\right] \label{eq:AngleAveragedFluxOpticallyThin}
\end{multline}    
In the $\Lya$ literature it is common to accommodate the alternative ``absorption-limited''(AL) case in which $\overline{\tau}_{\nu} > 0$, by defining the ``attenuation length,'' $r_0$, as $\exp[-\tau_{\nu}(z, z^{\prime})] \equiv \exp[-l_H(z, z^{\prime}) / r_0]$,
where $l_H$ is the proper distance between redshifts $z$ and $z^{\prime}$. Instead, we will adopt the neutral-medium approximation of Equation \ref{eq:TauNeutral} (i.e., $\xibar=0$), which permits the partially analytic solution
\begin{multline}
    \widehat{J}_{\nu,\mathrm{AL}}(z) = \frac{c}{4\pi} \frac{\hat{\upepsilon}_{\nu}(z)}{H(z)} (1 + z)^{9/2-(\alpha + \beta)} \\
     \times \exp\left[-\left(\frac{\mu}{\nu}\right)^3 (1 + z)^{3/2} \right] \mathcal{A}_{\nu}(\alpha, \beta, z, z_f) \label{eq:AngleAveragedFluxNeutral}
\end{multline}
with
\begin{equation}
    \mathcal{A}_{\nu} \equiv \int_{z^{\prime} = z}^{z^{\prime} = z_f} (1+z^{\prime})^{\alpha + \beta - 5/2} \exp\left[\left(\frac{\mu}{\nu}\right)^3 \frac{(1 + z)^3}{(1+z^{\prime})^{3/2}} \right] dz^{\prime} . \label{eq:Anu}
\end{equation}    
The function $A_{\nu}$ has analytic solutions (in the form of Exponential integrals) only for $\alpha + \beta = 3n/2$ where $n$ is a positive integer, which represents physically unrealistic scenarios.

\begin{figure}
\includegraphics[width=0.49\textwidth]{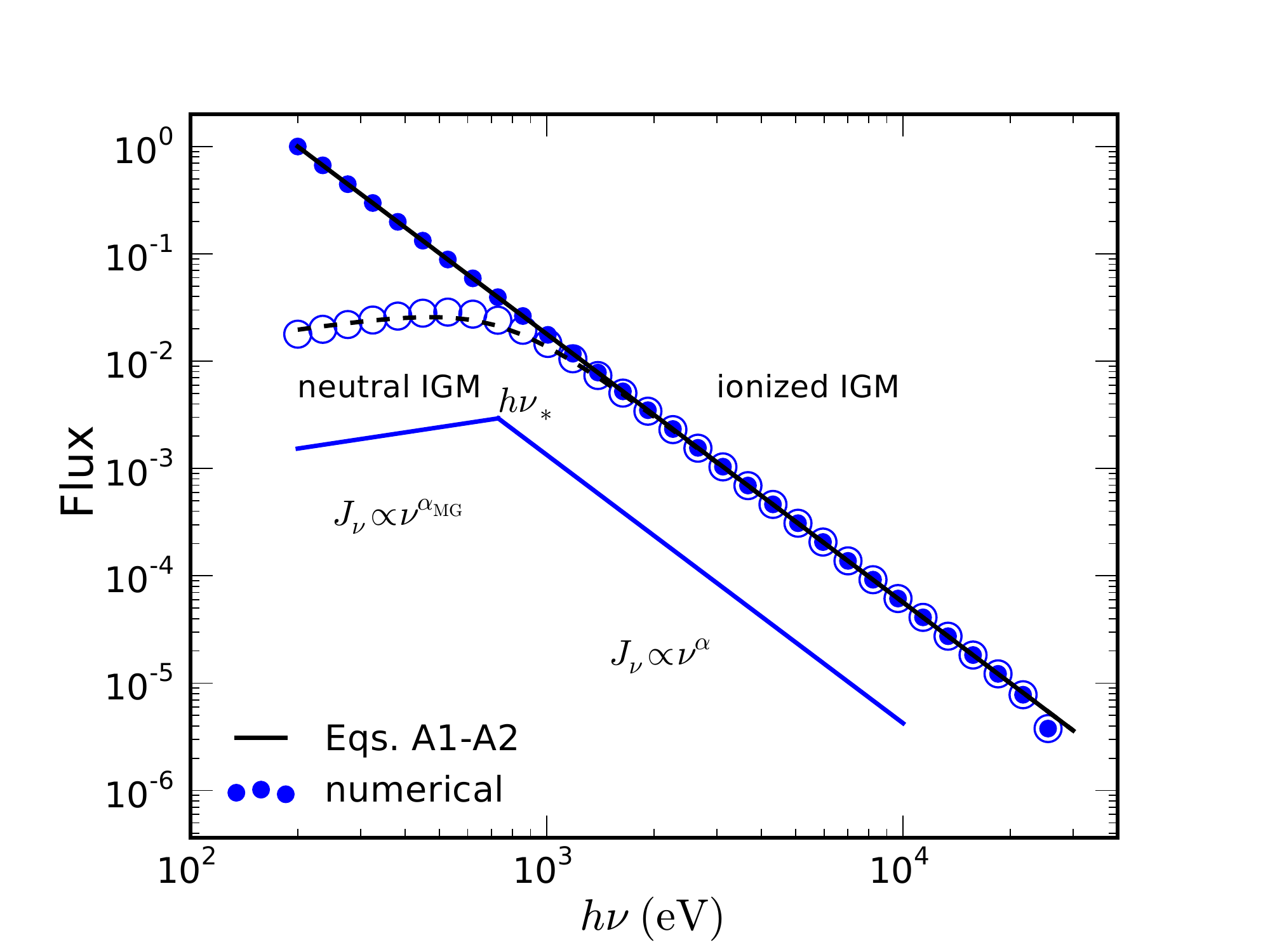}
\caption{Cosmic X-ray background spectrum at $z=20$ for $\alpha=-1.5$ and $\beta=-3$. Normalization of the $y$-axis can be scaled arbitrarily depending on the normalization of the emissivity. The deviation at high energies is due to the fact that the analytic solution is not truncated by $z_f$ or $E_{\max}$, meaning there are always higher energy photons redshifting to energies $h\nu \leq h\nu_{\max}$. The numerical solutions are computed with finite integration limits and truncated at $E_{\max}$, such that the emissivity at $h\nu > h\nu_{\max}$ is zero, resulting in no flux at $h\nu \geq h\nu_{\max}$. Elsewhere, the agreement is very good, with discrepancies arising solely due to the use of approximate bound-free photo-ionization cross sections in the analytic solution.}
\label{fig:test_cxrb}
\end{figure}

The metagalactic spectral index in this case works out to be
\begin{equation}
    \alphaMG \equiv \frac{d\log J_{\nu}}{d\log \nu} = \alpha + 3 \left(\frac{\mu}{\nu}\right)^3 (1+z)^{3/2} \left[1 - \mathcal{B}_{\nu} (1 + z)^{3/2} \right]
\end{equation}
where
\begin{equation}
    \mathcal{B}_{\nu} = \mathcal{A}_{\nu}^{-1} \int_z^{z_f} (1 + z^{\prime})^{\alpha + \beta - 4} \exp\left[\left(\frac{\mu}{\nu}\right)^{3} \frac{(1 + z)^{3}}{(1 + z^{\prime})^{3/2}} \right] dz^{\prime} .
\end{equation}
As $\nu \rightarrow \infty$, the second term vanishes, leaving the optically-thin limit, $\alphaMG = \alpha$. As $\nu \rightarrow 0$, $\mathcal{B}_{\nu} \rightarrow 0$, meaning $\alphaMG = \alpha + 3$. The ``break'' in the cosmic X-ray background spectrum occurs when $\alphaMG = 0$, corresponding to a photon energy of
\begin{equation}
    h\nu_{\ast} = h\mu (1 + z) \left\{\frac{3}{\alpha}\left[\mathcal{B}_{\nu_{\ast}} - (1 + z)^{-3/2}\right]\right\}^{1/3}
\end{equation}    
which must be solved iteratively. Solutions are presented in Figure \ref{fig:test_cxrb} for $\alpha=-1.5$, $\beta=-3$, $\enu(z_0) = 10^{-2}$ for $z_0 = 10$, $z_f = 15$, and show good agreement between analytic and numerical solutions. 

\end{document}

%% file: macros.tex
\newcommand*{\dt}[1]{%
  \accentset{\mbox{\large\bfseries .}}{#1}}

\newcommand{\kB}{k_{\text{B}}}

\newcommand{\Omnow}{\Omega_{\text{m},0}}
\newcommand{\Obnow}{\Omega_{\text{b},0}}

\newcommand{\Hofz}{H(z)}

\newcommand{\Tcmb}{T_{\gamma}}

\newcommand{\HI}{\text{H} {\textsc{i}}}

\newcommand{\HeI}{\text{He} {\textsc{i}}}
\newcommand{\HeII}{\text{He} {\textsc{ii}}}

\newcommand{\Lya}{\text{Ly-}\alpha}
\newcommand{\Lyn}{\text{Ly-}n}

\newcommand{\nmax}{n_{\text{max}}}

\newcommand{\frecn}{f_{\text{rec}}^{(n)}}

\newcommand{\nH}{n_{\text{H}}}

\newcommand{\nHbar}{\bar{n}_{\text{H}}^0}

\newcommand{\nbbar}{\bar{n}_{\text{b}}^0}
\newcommand{\rhobbar}{\bar{\rho}_{\text{b}}^0}

\newcommand{\NHI}{N_{\text{H } \textsc{i}}}

\newcommand{\Jhatnu}{\widehat{J}_{\nu}}

\newcommand{\ecomp}{\upepsilon_{\text{comp}}}

\newcommand{\eX}{\upepsilon_X}

\newcommand{\enu}{\hat{\upepsilon}_{\nu}}

\newcommand{\enuprime}{\hat{\upepsilon}_{\nu^{\prime}}}

\newcommand{\alphaMG}{\alpha_{\text{MG}}}

\newcommand{\taubar}{\overline{\tau}_{\nu}}

\newcommand{\xHII}{x_{\text{H } \textsc{ii}}}

\newcommand{\xHeII}{x_{\text{He } \textsc{ii}}}
\newcommand{\xHeIII}{x_{\text{He } \textsc{iii}}}

\newcommand{\xibar}{\overline{x}_i}

\newcommand{\nHII}{n_{\text{H } \textsc{ii}}}

\newcommand{\nHeII}{n_{\text{He } \textsc{ii}}}

\newcommand{\nel}{n_{\text{e}}}  
\newcommand{\ntot}{n_{\text{tot}}}

\newcommand{\acc}{\text{acc}}

\newcommand{\fedd}{f_{\text{edd}}}
\newcommand{\Mbh}{M_{\bullet}}

\newcommand{\fX}{f_X}

\newcommand{\fsc}{f_{\text{sc}}}

\newcommand{\fstar}{f_{\ast}}

\newcommand{\fbh}{f_{\bullet}}
\newcommand{\fcoll}{f_{\text{coll}}}

\newcommand{\rhobhdot}{\dt{\rho}_{\bullet}}

\newcommand{\rhostardot}{\dt{\rho}_{\ast}}

\newcommand{\Nion}{N_{\text{ion}}}

\newcommand{\fesc}{f_{\text{esc}}}
\newcommand{\Msun}{M_{\odot}}

\newcommand{\Tmin}{T_{\text{min}}}

\newcommand{\fion}{f_{\text{ion}}}

\newcommand{\nuHub}{\nu_{\mathrm{Hub}} }

\newcommand{\fheat}{f_{\text{heat}}}

\newcommand{\Lbol}{\mathcal{L}_{\text{bol}}}

\newcommand{\dTb}{\delta T_b}

\newcommand{\intensityunitsnumber}{\text{s}^{-1} \ \text{cm}^{-2} \ \mathrm{Hz}^{-1} \ \text{sr}^{-1}}

\newcommand{\coemissivityunitsnumber}{\text{s}^{-1} \ \mathrm{Hz}^{-1} \ \text{cMpc}^{-3}}
\newcommand{\coemissivityunitsenergy}{\text{erg} \ \text{s}^{-1} \ \mathrm{Hz}^{-1} \ \text{cMpc}^{-3}}

\newcommand{\cXunits}{\text{erg} \ \text{s}^{-1} \ (\Msun \ \text{yr})^{-1}}

\newcommand{\dprime}{\prime\prime}

%% file: ms.bbl
\begin{thebibliography}{98}
\expandafter\ifx\csname natexlab\endcsname\relax\def\natexlab#1{#1}\fi

\bibitem[{{Abel} {et~al}\mbox{.}(2002){Abel}, {Bryan}, \& {Norman}}]{Abel2002b}
{Abel} T., {Bryan} G.~L., {Norman} M.~L., 2002, Science, 295, 93

\bibitem[{{Bardeen}(1970)}]{Bardeen1970}
{Bardeen} J.~M., 1970, \nat, 226, 64

\bibitem[{Barkana \& Loeb(2001)}]{Barkana2001}
Barkana R., Loeb A., 2001, Physics Reports, 349, 125

\bibitem[{Barkana \& Loeb(2005)}]{Barkana2005}
Barkana R., Loeb A., 2005, \apj, 626, 1

\bibitem[{Basu-Zych {et~al}\mbox{.}(2013)Basu-Zych, Lehmer, Hornschemeier,
  Bouwens, Fragos, Oesch, Belczynski, Brandt, Kalogera, Luo, Miller, Mullaney,
  Tzanavaris, Xue, \& Zezas}]{BasuZych2013}
Basu-Zych A.~R. {et~al.}, 2013, \apj, 762, 45

\bibitem[{{Begelman} {et~al}\mbox{.}(2006){Begelman}, {Volonteri}, \&
  {Rees}}]{Begelman2006}
{Begelman} M.~C., {Volonteri} M., {Rees} M.~J., 2006, \mnras, 370, 289

\bibitem[{Belczynski {et~al}\mbox{.}(2008)Belczynski, Kalogera, Rasio, Taam,
  Zezas, Bulik, Maccarone, \& Ivanova}]{Belczynski2008}
Belczynski K. {et~al.}, 2008, \apjs, 174, 223

\bibitem[{Benson {et~al}\mbox{.}(2013)Benson, Venkatesan, \&
  Shull}]{Benson2013}
Benson A., Venkatesan A., Shull J.~M., 2013, \apj, 770, 76

\bibitem[{{Bromm} {et~al}\mbox{.}(1999){Bromm}, {Coppi}, \&
  {Larson}}]{Bromm1999}
{Bromm} V., {Coppi} P.~S., {Larson} R.~B., 1999, \apjl, 527, L5

\bibitem[{Bromm {et~al}\mbox{.}(2001)Bromm, Kudritzki, \& Loeb}]{Bromm2001}
Bromm V., Kudritzki R.~P., Loeb A., 2001, \apj, 552, 464

\bibitem[{{Bromm} {et~al}\mbox{.}(2009){Bromm}, {Yoshida}, {Hernquist}, \&
  {McKee}}]{Bromm2009}
{Bromm} V. {et~al.}, 2009, \nat, 459, 49

\bibitem[{Brorby {et~al}\mbox{.}(2014)Brorby, Kaaret, \&
  Prestwich}]{Brorby2014}
Brorby M., Kaaret P., Prestwich A., 2014, preprint (astro-ph/14043132)

\bibitem[{{Burns} {et~al}\mbox{.}(2012){Burns}, {Lazio}, {Bale}, {Bowman},
  {Bradley}, {Carilli}, {Furlanetto}, {Harker}, {Loeb}, \&
  {Pritchard}}]{Burns2012}
{Burns} J.~O. {et~al.}, 2012, Advances in Space Research, 49, 433

\bibitem[{{Chen} \& {Miralda-Escud{\'e}}(2004)}]{Chen2004}
{Chen} X., {Miralda-Escud{\'e}} J., 2004, \apj, 602, 1

\bibitem[{Chuzhoy {et~al}\mbox{.}(2006)Chuzhoy, Alvarez, \&
  Shapiro}]{Chuzhoy2006}
Chuzhoy L., Alvarez M.~A., Shapiro P.~R., 2006, The Astrophysical Journal, 648,
  L1

\bibitem[{Ciardi {et~al}\mbox{.}(2012)Ciardi, Bolton, \& Maselli}]{Ciardi2012}
Ciardi B., Bolton J.~S., Maselli A., 2012, \mnras

\bibitem[{Ciardi {et~al}\mbox{.}(2010)Ciardi, Salvaterra, \&
  Di~Matteo}]{Ciardi2010}
Ciardi B., Salvaterra R., Di~Matteo T., 2010, \mnras, 401, 2635

\bibitem[{Davies \& Furlanetto(2013)}]{Davies2013}
Davies F.~B., Furlanetto S.~R., 2013, \mnras

\bibitem[{{Dijkstra} {et~al}\mbox{.}(2012){Dijkstra}, {Gilfanov}, {Loeb}, \&
  {Sunyaev}}]{Dijkstra2012}
{Dijkstra} M. {et~al.}, 2012, \mnras, 421, 213

\bibitem[{Fabbiano(2006)}]{Fabbiano2006}
Fabbiano G., 2006, Annual Review of Astronomy and Astrophysics, 44, 323

\bibitem[{{Fialkov} {et~al}\mbox{.}(2014){Fialkov}, {Barkana}, \&
  {Visbal}}]{Fialkov2014}
{Fialkov} A., {Barkana} R., {Visbal} E., 2014, \nat, 506, 197

\bibitem[{{Field}(1958)}]{Field1958}
{Field} G.~B., 1958, Proceedings of the IRE, 46, 240

\bibitem[{Fragos {et~al}\mbox{.}(2013)Fragos, Lehmer, Naoz, Zezas, \&
  Basu-Zych}]{Fragos2013}
Fragos T. {et~al.}, 2013, \apj, 776, L31

\bibitem[{Friedrich {et~al}\mbox{.}(2012)Friedrich, Mellema, Iliev, \&
  Shapiro}]{Friedrich2012}
Friedrich M.~M. {et~al.}, 2012, Monthly Notices of the Royal Astronomical
  Society, 421, 2232

\bibitem[{Fukugita \& Kawasaki(1994)}]{Fukugita1994}
Fukugita M., Kawasaki M., 1994, \mnras, 269, 563

\bibitem[{Furlanetto(2006)}]{Furlanetto2006}
Furlanetto S.~R., 2006, \mnras, 371, 867

\bibitem[{Furlanetto \& Loeb(2004)}]{Furlanetto2004}
Furlanetto S.~R., Loeb A., 2004, \apj, 611, 642

\bibitem[{Furlanetto {et~al}\mbox{.}(2006)Furlanetto, Oh, \&
  Briggs}]{FurlanettoOhBriggs2006}
Furlanetto S.~R., Oh S.~P., Briggs F.~H., 2006, Physics Reports, 433, 181

\bibitem[{{Furlanetto} \& {Pritchard}(2006)}]{FurlanettoPritchard2006}
{Furlanetto} S.~R., {Pritchard} J.~R., 2006, \mnras, 372, 1093

\bibitem[{{Furlanetto} \& {Stoever}(2010)}]{Furlanetto2010}
{Furlanetto} S.~R., {Stoever} S.~J., 2010, \mnras, 404, 1869

\bibitem[{{Gilfanov} {et~al}\mbox{.}(2004){Gilfanov}, {Grimm}, \&
  {Sunyaev}}]{Gilfanov2004}
{Gilfanov} M., {Grimm} H.-J., {Sunyaev} R., 2004, \mnras, 347, L57

\bibitem[{Gnedin {et~al}\mbox{.}(2008)Gnedin, Kravtsov, \& Chen}]{Gnedin2008}
Gnedin N.~Y., Kravtsov A.~V., Chen H.~W., 2008, \apj

\bibitem[{Greenhill \& Bernardi(2012)}]{Greenhill2012}
Greenhill L.~J., Bernardi G., 2012, preprint (astro-ph/12011700)

\bibitem[{{Greiner} {et~al}\mbox{.}(2009){Greiner}, {Kr\u hler}, {Fynbo},
  {Rossi}, {Schwarz}, {Klose}, {Savaglio}, {Tanvir}, \&
  {McBreen}}]{Greiner2009}
{Greiner} J. {et~al.}, 2009, \apj, 693, 1610

\bibitem[{{Grimm} {et~al}\mbox{.}(2003){Grimm}, {Gilfanov}, \&
  {Sunyaev}}]{Grimm2003}
{Grimm} H.-J., {Gilfanov} M., {Sunyaev} R., 2003, \mnras, 339, 793

\bibitem[{{Haardt} \& {Madau}(1996)}]{Haardt1996}
{Haardt} F., {Madau} P., 1996, \apj, 461, 20

\bibitem[{{Haiman} {et~al}\mbox{.}(2000){Haiman}, {Abel}, \&
  {Rees}}]{Haiman2000}
{Haiman} Z., {Abel} T., {Rees} M.~J., 2000, \apj, 534, 11

\bibitem[{Haiman {et~al}\mbox{.}(1998)Haiman, Haiman, Loeb, \&
  Loeb}]{Haiman1998}
Haiman Z. {et~al.}, 1998, \apj, 503, 505

\bibitem[{Haiman {et~al}\mbox{.}(1997)Haiman, Rees, \& Loeb}]{Haiman1997}
Haiman Z., Rees M.~J., Loeb A., 1997, \apj

\bibitem[{{Harker} {et~al}\mbox{.}(2012){Harker}, {Pritchard}, {Burns}, \&
  {Bowman}}]{Harker2012}
{Harker} G.~J.~A. {et~al.}, 2012, \mnras, 419, 1070

\bibitem[{Hirata(2006)}]{Hirata2006}
Hirata C.~M., 2006, \mnras, 367, 259

\bibitem[{{Jeon} {et~al}\mbox{.}(2014){Jeon}, {Pawlik}, {Bromm}, \&
  {Milosavljevi{\'c}}}]{Jeon2014}
{Jeon} M. {et~al.}, 2014, \mnras, 440, 3778

\bibitem[{Kaaret(2014)}]{Kaaret2014}
Kaaret P., 2014, \mnrasl

\bibitem[{Kaaret {et~al}\mbox{.}(2011)Kaaret, Schmitt, \& Gorski}]{Kaaret2011}
Kaaret P., Schmitt J., Gorski M., 2011, \apj, 741, 10

\bibitem[{{King} \& {Kolb}(1999)}]{King1999}
{King} A.~R., {Kolb} U., 1999, \mnras, 305, 654

\bibitem[{{Komatsu} {et~al}\mbox{.}(2011){Komatsu}, {Smith}, {Dunkley},
  {Bennett}, {Gold}, {Hinshaw}, {Jarosik}, {Larson}, {Nolta}, {Page},
  {Spergel}, {Halpern}, {Hill}, {Kogut}, {Limon}, {Meyer}, {Odegard}, {Tucker},
  {Weiland}, {Wollack}, \& {Wright}}]{Komatsu2011}
{Komatsu} E. {et~al.}, 2011, \apjs, 192, 18

\bibitem[{{Kuhlen} \& {Madau}(2005)}]{Kuhlen2005}
{Kuhlen} M., {Madau} P., 2005, \mnras, 363, 1069

\bibitem[{{Lewis} {et~al}\mbox{.}(2000){Lewis}, {Challinor}, \&
  {Lasenby}}]{Lewis2000}
{Lewis} A., {Challinor} A., {Lasenby} A., 2000, \apj, 538, 473

\bibitem[{{Linden} {et~al}\mbox{.}(2010){Linden}, {Kalogera}, {Sepinsky},
  {Prestwich}, {Zezas}, \& {Gallagher}}]{Linden2010}
{Linden} T. {et~al.}, 2010, \apj, 725, 1984

\bibitem[{Madau {et~al}\mbox{.}(1997)Madau, Meiksin, \& Rees}]{Madau1997}
Madau P., Meiksin A., Rees M.~J., 1997, \apj, 475, 429

\bibitem[{Madau {et~al}\mbox{.}(2004)Madau, Rees, Volonteri, Haardt, \&
  Oh}]{Madau2004}
Madau P. {et~al.}, 2004, \apj, 604, 484

\bibitem[{Mapelli {et~al}\mbox{.}(2010)Mapelli, Ripamonti, Zampieri, Colpi, \&
  Bressan}]{Mapelli2010}
Mapelli M. {et~al.}, 2010, \mnras, 408, 234

\bibitem[{Mcquinn(2012)}]{McQuinn2012b}
Mcquinn M., 2012, \mnras, 426, 1349

\bibitem[{Meiksin \& White(2003)}]{Meiksin2003}
Meiksin A., White M., 2003, \mnras, 342, 1205

\bibitem[{Mesinger {et~al}\mbox{.}(2009)Mesinger, Bryan, \&
  Haiman}]{Mesinger2009a}
Mesinger A., Bryan G.~L., Haiman Z., 2009, \mnras, 399, 1650

\bibitem[{Mesinger {et~al}\mbox{.}(2013)Mesinger, Ferrara, \&
  Spiegel}]{Mesinger2013a}
Mesinger A., Ferrara A., Spiegel D.~S., 2013, \mnras

\bibitem[{Mesinger \& Furlanetto(2009)}]{Mesinger2009b}
Mesinger A., Furlanetto S., 2009, \mnras, 400, 1461

\bibitem[{{Mesinger} {et~al}\mbox{.}(2011){Mesinger}, {Furlanetto}, \&
  {Cen}}]{Mesinger2011}
{Mesinger} A., {Furlanetto} S., {Cen} R., 2011, \mnras, 411, 955

\bibitem[{Miller {et~al}\mbox{.}(2009)Miller, Cackett, \& Reis}]{Miller2009}
Miller J.~M., Cackett E.~M., Reis R.~C., 2009, \apj, 707, L77

\bibitem[{Mineo {et~al}\mbox{.}(2012)Mineo, Gilfanov, \& Sunyaev}]{Mineo2012b}
Mineo S., Gilfanov M., Sunyaev R., 2012, \mnras

\bibitem[{{Mineo} {et~al}\mbox{.}(2012){Mineo}, {Gilfanov}, \&
  {Sunyaev}}]{Mineo2012}
{Mineo} S., {Gilfanov} M., {Sunyaev} R., 2012, \mnras, 419, 2095

\bibitem[{{Mirabel} {et~al}\mbox{.}(2011){Mirabel}, {Dijkstra}, {Laurent},
  {Loeb}, \& {Pritchard}}]{Mirabel2011}
{Mirabel} I.~F. {et~al.}, 2011, \aap, 528, A149

\bibitem[{{Mirocha} {et~al}\mbox{.}(2013){Mirocha}, {Harker}, \&
  {Burns}}]{Mirocha2013}
{Mirocha} J., {Harker} G.~J.~A., {Burns} J.~O., 2013, \apj, 777, 118

\bibitem[{{Mirocha} {et~al}\mbox{.}(2012){Mirocha}, {Skory}, {Burns}, \&
  {Wise}}]{Mirocha2012}
{Mirocha} J. {et~al.}, 2012, \apj, 756, 94

\bibitem[{Mitsuda {et~al}\mbox{.}(1984)Mitsuda, Inoue, Koyama, Makishima,
  Matsuoka, Ogawara, Suzuki, Tanaka, Shibazaki, \& Hirano}]{Mitsuda1984}
Mitsuda K. {et~al.}, 1984, \pasj, 36, 741

\bibitem[{{Murray} {et~al}\mbox{.}(2013){Murray}, {Power}, \&
  {Robotham}}]{Murray2013}
{Murray} S.~G., {Power} C., {Robotham} A.~S.~G., 2013, Astronomy and Computing,
  3, 23

\bibitem[{Oh(2001)}]{Oh2001}
Oh S.~P., 2001, \apj, 553, 499

\bibitem[{{Pacucci} {et~al}\mbox{.}(2014){Pacucci}, {Mesinger}, {Mineo}, \&
  {Ferrara}}]{Pacucci2014}
{Pacucci} F. {et~al.}, 2014, eprint (astro-ph/14036125)

\bibitem[{Parsons {et~al}\mbox{.}(2014)Parsons, Liu, Aguirre, Ali, Bradley,
  Carilli, DeBoer, Dexter, Gugliucci, Jacobs, Klima, MacMahon, Manley, Moore,
  Pober, Stefan, \& Walbrugh}]{Parsons2014}
Parsons A.~R. {et~al.}, 2014, \apj, 788, 106

\bibitem[{Power {et~al}\mbox{.}(2013)Power, James, Combet, \& Wynn}]{Power2013}
Power C. {et~al.}, 2013, \apj, 764, 76

\bibitem[{Press \& Schechter(1974)}]{PressSchecter1974}
Press W.~H., Schechter P., 1974, \apj, 187, 425

\bibitem[{Prestwich {et~al}\mbox{.}(2013)Prestwich, Tsantaki, Zezas, Jackson,
  Roberts, Foltz, Linden, \& Kalogera}]{Prestwich2013}
Prestwich A.~H. {et~al.}, 2013, \apj, 769, 92

\bibitem[{Pritchard \& Furlanetto(2006)}]{Pritchard2006}
Pritchard J.~R., Furlanetto S.~R., 2006, \mnras, 367, 1057

\bibitem[{Pritchard \& Furlanetto(2007)}]{Pritchard2007}
Pritchard J.~R., Furlanetto S.~R., 2007, \mnras, 376, 1680

\bibitem[{{Pritchard} \& {Loeb}(2012)}]{Pritchard2012}
{Pritchard} J.~R., {Loeb} A., 2012, Reports on Progress in Physics, 75, 086901

\bibitem[{{Ranalli} {et~al}\mbox{.}(2003){Ranalli}, {Comastri}, \&
  {Setti}}]{Ranalli2003}
{Ranalli} P., {Comastri} A., {Setti} G., 2003, \aap, 399, 39

\bibitem[{Ricotti \& Ostriker(2004)}]{Ricotti2004}
Ricotti M., Ostriker J.~P., 2004, \mnras, 352, 547

\bibitem[{Ripamonti {et~al}\mbox{.}(2008)Ripamonti, Mapelli, \&
  Zaroubi}]{Ripamonti2008}
Ripamonti E., Mapelli M., Zaroubi S., 2008, \mnras, 387, 158

\bibitem[{Santos {et~al}\mbox{.}(2010)Santos, Ferramacho, Silva, Amblard, \&
  Cooray}]{Santos2010}
Santos M.~G. {et~al.}, 2010, \mnras, 406, 2421

\bibitem[{{Schaerer}(2002)}]{Schaerer2002}
{Schaerer} D., 2002, \aap, 382, 28

\bibitem[{Shakura \& Sunyaev(1973)}]{ShakuraSunyaev1973}
Shakura N.~I., Sunyaev R.~A., 1973, \aap, 24, 337

\bibitem[{{Shapiro} {et~al}\mbox{.}(1976){Shapiro}, {Lightman}, \&
  {Eardley}}]{Shapiro1976}
{Shapiro} S.~L., {Lightman} A.~P., {Eardley} D.~M., 1976, \apj, 204, 187

\bibitem[{{Shull} \& {van Steenberg}(1985)}]{Shull1985}
{Shull} J.~M., {van Steenberg} M.~E., 1985, \apj, 298, 268

\bibitem[{Steiner {et~al}\mbox{.}(2009)Steiner, Narayan, McClintock, \&
  Ebisawa}]{Steiner2009}
Steiner J.~F. {et~al.}, 2009, \pasp, 121, 1279

\bibitem[{{Tanaka} {et~al}\mbox{.}(2012){Tanaka}, {Perna}, \&
  {Haiman}}]{Tanaka2012}
{Tanaka} T., {Perna} R., {Haiman} Z., 2012, \mnras, 425, 2974

\bibitem[{Thomas \& Zaroubi(2008)}]{Thomas2008}
Thomas R.~M., Zaroubi S., 2008, \mnras, 384, 1080

\bibitem[{{Totani} {et~al}\mbox{.}(2006){Totani}, {Kawai}, {Kosugi}, {Aoki},
  {Yamada}, {Iye}, {Ohta}, \& {Hattori}}]{Totani2006}
{Totani} T. {et~al.}, 2006, \pasj, 58, 485

\bibitem[{Tumlinson {et~al}\mbox{.}(2003)Tumlinson, Shull, \&
  Venkatesan}]{Tumlinson2003}
Tumlinson J., Shull J.~M., Venkatesan A., 2003, \apj, 584, 608

\bibitem[{Venkatesan \& Benson(2011)}]{Venkatesan2011}
Venkatesan A., Benson A., 2011, \mnras, 417, 2264

\bibitem[{Venkatesan {et~al}\mbox{.}(2001)Venkatesan, Venkatesan, Giroux,
  Giroux, Shull, \& Shull}]{Venkatesan2001}
Venkatesan A. {et~al.}, 2001, \apj, 563, 1

\bibitem[{Verner \& Ferland(1996)}]{Verner1996}
Verner D.~A., Ferland G.~J., 1996, \apjs, 103, 467

\bibitem[{{Volonteri} {et~al}\mbox{.}(2005){Volonteri}, {Madau}, {Quataert}, \&
  {Rees}}]{Volonteri2005}
{Volonteri} M. {et~al.}, 2005, \apj, 620, 69

\bibitem[{{Voytek} {et~al}\mbox{.}(2014){Voytek}, {Natarajan}, {J{\'a}uregui
  Garc{\'{\i}}a}, {Peterson}, \& {L{\'o}pez-Cruz}}]{Voytek2014}
{Voytek} T.~C. {et~al.}, 2014, \apjl, 782, L9

\bibitem[{Wise \& Cen(2009)}]{Wise2009}
Wise J.~H., Cen R., 2009, \apj, 693, 984

\bibitem[{Wolcott-Green \& Haiman(2012)}]{WolcottGreen2012}
Wolcott-Green J., Haiman Z., 2012, \mnrasl, 425, L51

\bibitem[{{Wouthuysen}(1952)}]{Wouthuysen1952}
{Wouthuysen} S.~A., 1952, \aj, 57, 31

\bibitem[{Wyithe \& Loeb(2003)}]{Wyithe2003}
Wyithe J. S.~B., Loeb A., 2003, \apj, 586, 693

\bibitem[{Zygelman(2005)}]{Zygelman2005}
Zygelman B., 2005, \apj, 622, 1356

\end{thebibliography}
